\documentclass[aps,epsf,showpacs,usenatbib]{iopart}
\pdfoutput=1
\usepackage{graphics}
\usepackage{graphicx}
\usepackage{epsfig}
\usepackage{times}
\usepackage{subfigure}
\usepackage{psfrag}
\usepackage{amssymb}
\usepackage{multirow}

\def\reff@jnl#1{{\rm#1\/}}
\def\aj{\reff@jnl{AJ}}                 
\def\araa{\reff@jnl{ARA\&A}}           
\def\apj{\reff@jnl{ApJ}}               
\def\apjl{\reff@jnl{ApJ}}              
\def\apjs{\reff@jnl{ApJS}}             
\def\ao{\reff@jnl{Appl.Optics}}        
\def\apss{\reff@jnl{Ap\&SS}}           
\def\aap{\reff@jnl{A\&A}}              
\def\aapr{\reff@jnl{A\&A~Rev.}}        
\def\aaps{\reff@jnl{A\&AS}}            
\def\azh{\reff@jnl{AZh}}               
\def\baas{\reff@jnl{BAAS}}             
\def\jrasc{\reff@jnl{JRASC}}           
\def\memras{\reff@jnl{MmRAS}}          
\def\mnras{\reff@jnl{MNRAS}}           
\def\pra{\reff@jnl{Phys.Rev.A}}        
\def\prb{\reff@jnl{Phys.Rev.B}}        
\def\prc{\reff@jnl{Phys.Rev.C}}        
\def\prd{\reff@jnl{Phys.Rev.D}}        
\def\prl{\reff@jnl{Phys.Rev.Lett}}     
\def\pasp{\reff@jnl{PASP}}             
\def\pasj{\reff@jnl{PASJ}}             
\def\qjras{\reff@jnl{QJRAS}}           
\def\skytel{\reff@jnl{S\&T}}           
\def\solphys{\reff@jnl{Solar~Phys.}}   
\def\sovast{\reff@jnl{Soviet~Ast.}}    
\def\ssr{\reff@jnl{Space~Sci.Rev.}}    
\def\zap{\reff@jnl{ZAp}}               
\def\nat{\reff@jnl{Nature}}            
\def\cqg{\reff@jnl{Class. Quantum Grav.}}    
\newcommand{\MN}{{\sc MultiNest}}

\begin{document}
\input epsf.tex

\title[Classifying gravitational wave bursts using Bayesian evidence]{Classifying LISA gravitational wave burst signals using Bayesian evidence}
\author{Farhan Feroz$^{1}$, Jonathan R. Gair$^{2}$\footnote[3]{jgair@ast.cam.ac.uk}, Philip Graff$^{1}$, Michael P Hobson$^{1}$ and Anthony Lasenby$^{3,1}$}
\address{$^{1}$Astrophysics Group, Cavendish Laboratory, JJ Thomson Avenue, Cambridge CB3 0HE, UK}
\address{$^{2}$Institute of Astronomy, Madingley Road, Cambridge CB3 0HA, UK} 
\address{$^{3}$Kavli Institute for Cosmology, Madingley Road, Cambridge CB3 0HA, UK}

\vspace{1cm}
\begin{abstract}
We consider the problem of characterisation of burst sources detected with the Laser Interferometer Space Antenna (LISA) using the multi-modal nested sampling algorithm, {\sc MultiNest}. We use {\sc MultiNest} as a tool to search for modelled bursts from cosmic string cusps, and compute the Bayesian evidence associated with the cosmic string model. As an alternative burst model, we consider sine-Gaussian burst signals, and show how the evidence ratio can be used to choose between these two alternatives. We present results from an application of {\sc MultiNest} to the last round of the Mock LISA Data Challenge, in which we were able to successfully detect and characterise all three of the cosmic string burst sources present in the release data set. We also present results of independent trials and show that {\sc MultiNest} can detect cosmic string signals with signal-to-noise ratio (SNR) as low as $\sim 7$ and sine-Gaussian signals with SNR as low as $\sim8$. In both cases, we show that the threshold at which the sources become detectable coincides with the SNR at which the evidence ratio begins to favour the correct model over the alternative.
\end{abstract}
\pacs{04.25.Nx, 04.30.Db, 04.80.Cc}

\maketitle

\section{Introduction}
The development of data analysis algorithms for gravitational wave detectors on the ground and for the proposed space-based gravitational wave detector, the Laser Interferometer Space Antenna (LISA)~\cite{lisa}, is an active area of current research. LISA data analysis activities are being encouraged by the Mock LISA Data Challenges~\cite{mldc} (MLDCs). Prior to the most recent round, round 3, which finished in April 2009, the MLDCs had included data sets containing individual and multiple white-dwarf binaries, a realisation of the whole galaxy of compact binaries, single and multiple non-spinning supermassive black hole (SMBH) binaries, either isolated or on top of a galactic confusion background and isolated extreme-mass-ratio inspiral (EMRI) sources in purely instrumental noise. Round 3 included a realisation of the galactic binaries that included chirping systems, a single data set containing multiple overlapping EMRIs and data sets containing three new types of source --- inspirals of spinning supermassive black hole binaries, bursts from cosmic string cusps and a stochastic background of gravitational radiation~\cite{mldc}. Several of these sources, including the EMRI signals, the spinning black hole binaries and the cosmic string bursts, have highly multi-modal likelihood surfaces which can cause problems for algorithms such as Markov Chain Monte Carlo (MCMC), as these may become stuck in secondary maxima rather than finding the primary mode~\cite{EtfAG,neilemri,BBGPa,BBGPb}. Recently~\cite{MNnospin} we applied a new algorithm to the problem of gravitational wave data analysis, {\sc MultiNest}~\cite{feroz08,multinest}, which is a nested sampling algorithm, optimized for problems with highly multi-modal posteriors. We demonstrated that {\sc MultiNest} could be used as a search tool and to recover posterior probability distributions for the case of data sets containing multiple signals from non-spinning SMBH binary inspirals. We used the algorithm as a search tool in MLDC round 3 to analyse two of the data sets --- the spinning SMBH inspirals (challenge 3.2) and the cosmic string bursts (challenge 3.4). We will discuss the second of these applications in this paper.

The nested sampling algorithm~\cite{Skilling04} was developed as a tool for evaluating the Bayesian evidence. It employs a set of live points, each of which represents a particular set of parameters in the multi-dimensional search space. These points move as the algorithm progresses, and climb together through nested contours of increasing likelihood.  At each step the algorithm works by finding a point of higher likelihood than the lowest likelihood point in the live point set and then replacing the lowest likelihood point with the new point. The difficulty is to sample, efficiently, new points of higher likelihood from the remaining prior volume. {\sc MultiNest} achieves this using an ellipsoidal rejection sampling scheme
. It has demonstrated orders of magnitude improvement over standard methods in several applications in cosmology and particle physics~\cite{2008arXiv08100781F,2008arXiv08111199F,Feroz:2008wr,2008JHEP12024T}. It also proved to be very effective in a gravitational wave context for the test case mentioned above~\cite{MNnospin}.

{\sc MultiNest} can be used as a search tool and returns the posterior probability distributions as a by-product. We employed the algorithm successfully to analyse MLDC challenge 3.4, accurately finding all three of the cosmic string burst sources present in the data set. This will be discussed in more detail in Section~\ref{sec:mldcres}. The cosmic string cusp is a special type of burst source, since the waveform is known and can be modelled, so a matched filtering search is possible and that is what {\sc MultiNest} does in computing the likelihoods across the parameter space. However, there may be bursts of gravitational waves in the LISA data stream from other sources and the question arises as to whether we would be able to detect them and if we can distinguish cosmic string bursts from bursts due to these other sources. The evidence that {\sc MultiNest} computes is one tool that can be used to address model selection. Evidence has been used for gravitational wave model selection to test the theory of relativity in ground based observations~\cite{Veitch:2008wd}, and, in a LISA context, to distinguish between an empty data set and one containing a signal~\cite{littenberg09}. Calculation of an evidence ratio requires a second model for the burst as an alternative to compare against. We chose to use a sine-Gaussian waveform as a generic alternative burst model, as this is one of the burst models commonly used in LIGO data analysis (see for instance~\cite{LSCUsingSG}). We find that the evidence is a powerful tool for characterising bursts --- for the majority of detectable bursts, the evidence ratio strongly favours the true model over the alternative. While the sine-Gaussian model does not necessarily well describe all possible un-modelled bursts, these results suggest that the evidence can be used to correctly identify any cosmic string bursts that are present in the LISA data stream.

This paper is organised as follows. In Section~\ref{sec:methods} we describe the methods that we employ in this analysis ---  we describe Bayesian inference, nested sampling, the {\sc MultiNest} algorithm, the two burst waveform models we use and the detector noise spectral density. In Section~\ref{sec:res} we present our results, including a description of our methods and results for the analysis of MLDC challenge 3.4 (in Section~\ref{sec:mldcres}); an estimate of the signal-to-noise ratio (SNR) required for detection of cosmic string and sine-Gaussian bursts using the {\sc MultiNest} algorithm; and a calculation of the evidence ratio of the two models for data sets containing each type of source. We finish in Section~\ref{sec:discuss} with a summary and discussion of directions for further research.

\section{Methodology}\label{sec:methods}
\subsection{Bayesian Inference}\label{sec:bayesian}
Bayesian inference methods provide a consistent approach to the estimation of a set of parameters~$\mathbf{\Theta}$ in a model (or hypothesis) $H$ for the data $\mathbf{D}$. Bayes' theorem states that
\begin{equation} \Pr(\mathbf{\Theta}|\mathbf{D}, H) =
\frac{\Pr(\mathbf{D}|\,\mathbf{\Theta},H)\Pr(\mathbf{\Theta}|H)}{\Pr(\mathbf{D}|H)},
\label{eq:bayes}
\end{equation}
where $\Pr(\mathbf{\Theta}|\mathbf{D}, H) \equiv P(\mathbf{\Theta})$ is the posterior probability distribution of the parameters, $\Pr(\mathbf{D}|\mathbf{\Theta}, H) \equiv \mathcal{L}(\mathbf{\Theta})$ is
the likelihood, $\Pr(\mathbf{\Theta}|H) \equiv \pi(\mathbf{\Theta})$ is the prior distribution, and $\Pr(\mathbf{D}|H) \equiv \mathcal{Z}$ is the Bayesian evidence.

Bayesian evidence is the factor required to normalise the posterior over~$\mathbf{\Theta}$:
\begin{equation}
\mathcal{Z} =
\int{\mathcal{L}(\mathbf{\Theta})\pi(\mathbf{\Theta})}d^N\mathbf{\Theta},
\label{eq:Z}
\end{equation}
where $N$ is the dimensionality of the parameter space. Since the Bayesian evidence is independent of the parameter values~$\mathbf{\Theta}$, it is usually ignored in parameter estimation problems and the
posterior inferences are obtained by exploring the un--normalized posterior using standard MCMC sampling methods.

Bayesian parameter estimation has been used quite extensively in a variety of astronomical applications, including gravitational wave astronomy, although standard MCMC methods, such as the basic Metropolis--Hastings algorithm or the Hamiltonian sampling technique (see e.g.~\cite{MacKay}), can experience problems in sampling efficiently from a multi--modal posterior distribution or one with large (curving) degeneracies between parameters. Moreover, MCMC methods often require careful tuning of the proposal distribution to sample efficiently, and testing for convergence can be problematic.

In order to select between two models $H_{0}$ and $H_{1}$ one needs to compare their respective posterior probabilities given the observed data set $\mathbf{D}$, as follows:
\begin{equation}
\frac{\Pr(H_{1}|\mathbf{D})}{\Pr(H_{0}|\mathbf{D})}
=\frac{\Pr(\mathbf{D}|H_{1})\Pr(H_{1})}{\Pr(\mathbf{D}|H_{0})\Pr(H_{0})}
=\frac{\mathcal{Z}_1}{\mathcal{Z}_0}\frac{\Pr(H_{1})}{\Pr(H_{0})},
\label{eq:model_select}
\end{equation}
where $\Pr(H_{1})/\Pr(H_{0})$ is the prior probability ratio for the two models, which can often be set to unity but occasionally requires further consideration (see, for example,
\cite{2008arXiv08100781F,2008arXiv08111199F} for some examples where the prior probability ratio should not be set to unity). It can be seen from Eq.~(\ref{eq:model_select}) that the Bayesian evidence plays a
central role in Bayesian model selection. As the average of likelihood over the prior, the evidence automatically implements Occam's razor: a simpler theory which agrees well enough with the empirical
evidence is preferred. A more complicated theory will only have a higher evidence if it is significantly better at explaining the data than a simpler theory.

Unfortunately, evaluation of Bayesian evidence involves the multidimensional integral (Eq.~(\ref{eq:Z})) and thus presents a challenging numerical task. Standard techniques like thermodynamic integration
\cite{Ruanaidh} are extremely computationally expensive which makes evidence evaluation typically at least an order of magnitude more costly than parameter estimation. Some fast approximate methods have been
used for evidence evaluation, such as treating the posterior as a multivariate Gaussian centred at its peak (see, for example, \cite{Hobson02}), but this approximation is clearly a poor one for highly non-Gaussian and
multi--modal posteriors. Various alternative information criteria for model selection are discussed in \cite{Liddle07}, but the evidence remains the preferred method.

\subsection{Nested Sampling and the {\sc MultiNest} Algorithm}\label{sec:multinest}

\begin{figure}
\begin{center}
\subfigure[]{\includegraphics[width=0.3\columnwidth]{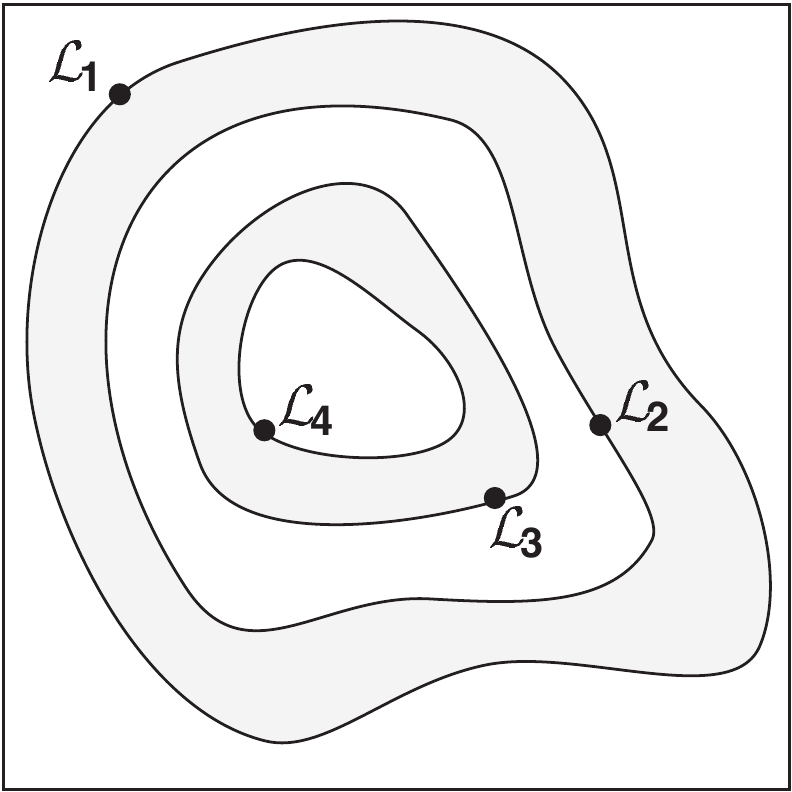}}\hspace{0.5cm}
\subfigure[]{\includegraphics[width=0.3\columnwidth]{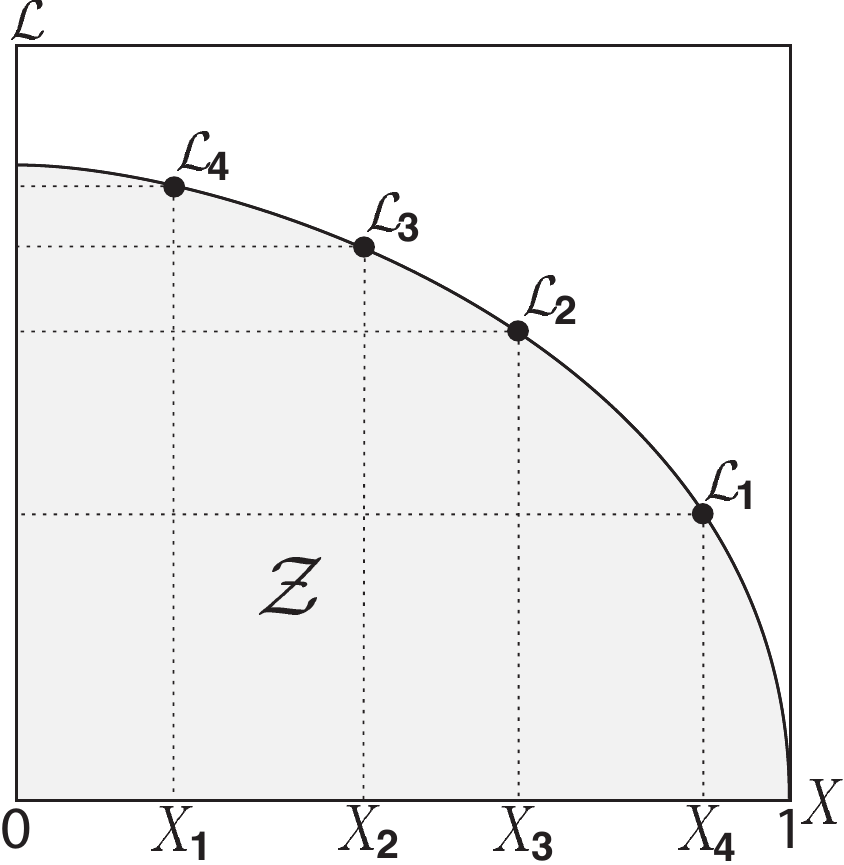}}
\caption{Cartoon illustrating (a) the posterior of a two dimensional problem; and (b) the transformed $\mathcal{L}(X)$  function where the prior volumes $X_{i}$ are associated with each likelihood
$\mathcal{L}_{i}$.}
\label{fig:NS}
\end{center}
\end{figure}

Nested sampling~\cite{Skilling04} is a Monte Carlo method targetted at the efficient calculation of the evidence, but also produces posterior inferences as a by-product. It calculates the evidence by
transforming the multi--dimensional evidence integral into a one--dimensional integral that is easy to evaluate numerically. This is accomplished by defining the prior volume $X$ as $dX =
\pi(\mathbf{\Theta})d^D \mathbf{\Theta}$, so that
\begin{equation}
X(\lambda) = \int_{\mathcal{L}\left(\mathbf{\Theta}\right) > \lambda} \pi(\mathbf{\Theta}) d^N\mathbf{\Theta},
\label{eq:Xdef}
\end{equation}
where the integral extends over the region(s) of parameter space contained within the iso-likelihood contour $\mathcal{L}(\mathbf{\Theta}) = \lambda$. The evidence integral, Eq.~(\ref{eq:Z}), can then be
written as
\begin{equation}
\mathcal{Z}=\int_{0}^{1}{\mathcal{L}(X)}dX,
\label{eq:nested}
\end{equation}
where $\mathcal{L}(X)$, the inverse of Eq.~(\ref{eq:Xdef}), is a  monotonically decreasing function of $X$.  Thus, if one can evaluate the likelihoods $\mathcal{L}_{i}=\mathcal{L}(X_{i})$, where $X_{i}$ is a
sequence of decreasing values,
\begin{equation}
0<X_{M}<\cdots <X_{2}<X_{1}< X_{0}=1,
\end{equation}
as shown schematically in Fig.~\ref{fig:NS}, the evidence can be approximated numerically using standard quadrature methods as a weighted sum
\begin{equation}
\mathcal{Z}={\textstyle {\displaystyle \sum_{i=1}^{M}}\mathcal{L}_{i}w_{i}},
\label{eq:NS_sum}
\end{equation}
where the weights $w_{i}$ for the simple trapezium rule are given by $w_{i}=\frac{1}{2}(X_{i-1}-X_{i+1})$. An example of a posterior in two dimensions and its associated function $\mathcal{L}(X)$ is shown in
Fig.~\ref{fig:NS}.

%

\subsubsection{Evidence Evaluation}\label{app:nested:evidence}

The summation in Eq.~(\ref{eq:NS_sum}) is performed as follows. The iteration counter is first set to~$i=0$ and $N$ `active' (or `live') samples are drawn from the full prior $\pi(\mathbf{\Theta})$, so the
initial prior volume is $X_{0} = 1$. The samples are then sorted in order of their likelihood and the smallest (with likelihood $\mathcal{L}_{0}$) is removed from the active set (hence becoming `inactive')
and replaced by a point drawn from the prior subject to the constraint that the point has a likelihood $\mathcal{L}>\mathcal{L}_{0}$. The corresponding prior volume contained within the iso-likelihood contour associated with the new live point will be a random variable given by $X_{1} = t_{1} X_{0}$, where $t_{1}$ follows the distribution $\Pr(t) = Nt^{N-1}$ (i.e., the probability distribution for the largest of $N$ samples drawn uniformly from the interval $[0,1]$). At each subsequent iteration $i$, the removal of the lowest likelihood point $\mathcal{L}_{i}$ in the active set, the drawing of a replacement with $\mathcal{L} > \mathcal{L}_{i}$ and the reduction of the corresponding prior volume $X_{i}=t_{i} X_{i-1}$ are repeated, until the entire prior volume has been traversed. The algorithm thus travels through nested shells of likelihood as the prior volume is reduced. The mean and standard deviation of $\log t$, which dominates the geometrical exploration, are: 
\begin{equation}
E[\log t] = -1/N, \quad \sigma[\log t] = 1/N.
\end{equation}
Since each value of $\log t$ is independent, after $i$ iterations the prior volume will shrink down such that $\log X_{i} \approx -(i\pm\sqrt{i})/N$. Thus, one takes $X_{i} = \exp(-i/N)$.

\subsubsection{Stopping Criterion}\label{nested:stopping}

The nested sampling algorithm is terminated when the evidence has been computed to a pre-specified precision. The evidence that could be contributed by the remaining live points is estimated as $\Delta{\mathcal{Z}}_{\rm i} = \mathcal{L}_{\rm max}X_{\rm i}$, where ${\cal L}_{\rm max}$ is the maximum-likelihood value of the remaining live points, and $X_i$ is the remaining prior volume. The algorithm terminates when $\Delta{\mathcal{Z}}_{\rm i}$ is less than a user-defined value (we use $0.5$ in log-evidence).

\subsubsection{Posterior Inferences}\label{nested:posterior}

Once the evidence~$\mathcal{Z}$ is found, posterior inferences can be easily  generated using the final live points and the full sequence of discarded points from the nested sampling  process, i.e., the points with the lowest
likelihood value at each iteration~$i$ of  the algorithm. Each such point is simply assigned the probability weight  \begin{equation} p_{i}=\frac{\mathcal{L}_{i}w_{i}}{\mathcal{Z}}.\label{eq:12}
\end{equation} These samples can then be used to calculate inferences of posterior parameters such as  means, standard deviations, covariances and so on, or to construct marginalised posterior distributions.

\subsubsection{{\sc MultiNest} Algorithm}\label{sec:method:bayesian:multinest}

%
%
The most challenging task in implementing nested sampling is to draw samples from the prior within the hard constraint $\mathcal{L}> \mathcal{L}_i$ at each iteration $i$. The {\sc MultiNest} algorithm
\cite{feroz08,multinest} tackles this problem through an ellipsoidal rejection sampling scheme. The  live point set is enclosed within a set of (possibly overlapping)  ellipsoids and a new point is then drawn
uniformly from the region enclosed by these ellipsoids. The ellipsoidal decomposition of the live point set is chosen to minimize the sum of volumes of the ellipsoids. The ellipsoidal decomposition is well suited to dealing with posteriors that have curving degeneracies, and allows mode identification in multi-modal posteriors. If there are subsets of the ellipsoid set that do not overlap with the remaining ellipsoids, these are identified as a distinct mode and subsequently evolved independently. The {\sc MultiNest} algorithm has proven very useful for tackling inference problems in cosmology and particle physics~\cite{2008arXiv08100781F,2008arXiv08111199F,Feroz:2008wr,2008JHEP12024T}, typically showing two orders of magnitude improvement in efficiency over conventional techniques. More recently, it has been shown to perform well as a search tool for gravitational wave data analysis~\cite{MNnospin}.

\subsection{Burst waveform models}
\subsubsection{Cosmic Strings}
Gravitational waves can be generated by cosmic strings through the formation of cusps, where portions of the string are traveling at nearly the speed of light~\cite{CScusps1,CScusps2}. Such radiation is highly beamed and, when viewed along the emission axis, is linearly polarized and takes a simple power-law form, $h(t) \propto |t-t_c|^{1/3}$~\cite{CScusps1}. When viewed slightly off-axis, the waveform is still approximately linearly polarized but the cusp spectrum is rounded off and decays rapidly for frequencies above $f_{\rm max} \sim 2/(\alpha^3 L)$, where $\alpha$ is the viewing angle and $L$ is the dimension of the feature generating the cusp~\cite{CScusps1,CornishCS}. The particular model for the frequency domain waveform adopted in the MLDC is given by~\cite{mldc}
\begin{equation}
|h(f)| = \left\{\begin{array}{ll}\mathcal{A} f^{-4/3}& f < f_{\rm max} \\ \mathcal{A} f^{-4/3} \textrm{exp}\left(1-\frac{f}{f_{max}}\right)& f > f_{\rm max}\end{array} \right. .
\end{equation}
In addition, the MLDC waveforms include a fourth-order Butterworth filter to mitigate dynamic-range issues associated with inverse Fourier transforms. 
We adopt the same ansatz as the MLDC, namely that the Fourier domain waveform amplitude is given by
\begin{eqnarray}
|h_+ | & = & \left\{\begin{array}{ll}\mathcal{A} f^{-4/3} \left(1+\left(\frac{f_{low}}{f}\right)^2\right)^{-4}& f < f_{\rm max} \\ \mathcal{A} f^{-4/3} \left(1+\left(\frac{f_{low}}{f}\right)^2\right)^{-4} \textrm{exp}\left(1-\frac{f}{f_{max}}\right)& f > f_{\rm max}\end{array} \right.\nonumber \\
| h_{\times} | & = & 0
\label{eq:cuspwave}
\end{eqnarray}
and the phase by exp($2 \pi \imath f t_c$), where $t_c$ is the burst time.

\subsubsection{Sine Gaussians}
A sine-Gaussian waveform is centred on a particular frequency, and has exponentially suppressed power at nearby frequencies. We choose to consider a lineraly polarised sine-Gaussian, for which the waveform magnitudes in the frequency domain are given by
\begin{equation}
| h_+ | = \frac{A}{2} \sqrt{\frac{Q^2}{2\pi f_c^2}} \times \textrm{exp}\left( -\frac{Q^2}{2} \left( \frac{f-f_c}{f_c} \right)^2 \right), \;\;\; | h_{\times} |= 0
\label{eq:sineGaussian}
\end{equation}
where $A$ is the dimensionless amplitude, $f_c$ is the frequency of the sinusoidal oscillation, and $Q$ is the width. The phase of the wave is again exp($2\pi \imath f t_c$), where $t_c$ is the burst time. In the time-domain, the sine-Gaussian is a small burst ``packet'' of particular frequency, $f_c$, with the number of cycles in the burst determined by $Q$.

\subsection{Detector model}
To include the LISA response we made use of the static LISA model as described in~\cite{statLISAresp}. This approximation is valid for burst sources, as LISA does not move significantly over the typically short duration of the bursts, which is of the order of $1000$s. The static LISA response model was also adopted for cosmic string bursts in~\cite{CornishCS}. 

Three optimal detection time-delay interferometry (TDI) channels, $A$, $E$ and $T$, can be constructed from the LISA data stream~\cite{tdi}. The noise is uncorrelated within and across these channels. For the search of the MLDC data, the $A$, $E$ and $T$ channels were constructed as linear combinations of the three Michelson channels, $X$, $Y$ and $Z$, that were provided in the data release. The power spectral densities of the noise in the three channels are given by
\begin{eqnarray}
\fl S_A(f) = S_E(f) &=& 16 \sin^2(2\pi f t_L) \left(2\left(1+\cos(2\pi f t_L)+\cos^2(2\pi f t_L)\right)S_{\rm pm}(f) \right. \nonumber \\ && \left.+ \left(1+\cos(2\pi f t_L)/2\right)S_{\rm sn}f^2\right)\label{sha} \\
\fl S_T(f) &=& 16 \sin^2(2\pi f t_L) \left(2\left(1-2\cos(2\pi f t_L)+\cos^2(2\pi f t_L)\right)S_{\rm pm}(f)  \right. \nonumber \\ && \left.+ \left(1-\cos(2\pi f t_L)\right)S_{\rm sn}f^2\right)\label{sht} \\
\fl S_{\rm pm}(f) &=&  \left(1+\left(\frac{10^{-4}{\rm Hz}}{f}\right)^2\right)\frac{S_{\rm acc}}{f^2}\nonumber \\
\label{specdens}
\end{eqnarray}
where $t_L=16.678$s is the light travel time along one arm of the LISA constellation, $S_{\rm acc}=2.5\times10^{-48}$Hz$^{-1}$ is the proof mass acceleration noise and $S_{\rm sn}=1.8\times10^{-37}$Hz$^{-1}$ is the shot noise.

The LISA data stream will also contain a confusion noise foreground from unresolved white dwarf binaries in our galaxy. These were not included in the noise model for the MLDC round 3.4, and so we have ignored the confusion noise contribution in the current work. In practice, we found that the $T$ channel was too noisy to be used in the MLDC search, and so we used the $A$ and $E$ channels only for data analysis.

\subsection{Likelihood evaluation}
The space of gravitational waveform signals possesses a natural scalar product~\cite{helst,owen}
\begin{equation}\label{eqn:scalarprod}
\left<h\left|s\right.\right> =2\int_{0}^{\infty}\frac{df}{S_{n}(f)}\,\left[ \tilde{h}(f)\tilde{s}^{*}(f) +  \tilde{h}^{*}(f)\tilde{s}(f) \right],
\label{eq:scalarprod}
\end{equation}
where 
\begin{equation}
\tilde{h}(f) = \int_{-\infty}^{\infty}\, dt\, h(t)e^{2\pi\imath ft}
\end{equation}
is the Fourier transform of the time domain waveform $h(t)$.  The quantity $S_{n}(f)$ is the one-sided noise spectral density of the detector. For a family of sources with waveforms~$h(t;{\vec{\lambda}})$
that depend on parameters~$\vec{\lambda}$, the output of the detector, $s(t) = h(t;\vec{\lambda_0}) + n(t)$, consists of the true signal~$h(t;\vec{\lambda_0})$ and a particular realisation of the noise,
$n(t)$. Assuming that the noise is stationary and Gaussian, the logarithm of the likelihood that the parameter values are given by~$\vec{\lambda}$ is
\begin{equation}
\log{\mathcal L}\left(\vec{\lambda}\right) = C-\left<s-h\left(\vec{\lambda}\right)\left|s-h\left(\vec{\lambda}\right)\right.\right>/2,  
\end{equation}
and it is this log-likelihood that is evaluated at each point in the search. The constant, $C$, depends on the dimensionality of the search space, but its value is not important as we are only interested in the relative likelihoods of different points. As mentioned earlier, the LISA data stream has several independent data channels. The total multi-channel likelihood is obtained by summing the scalar product~(\ref{eqn:scalarprod}) over the various channels (in our work we use only the two channels, $A$ and $E$).

\subsection{Priors}\label{sec:MLDCpriors}
The parameter space over which to search for signals must also be specified. For the searches with the cosmic string model, we used uniform priors for each parameter that covered the signal space from which the MLDC sources were drawn. These correspond to the following ranges:
\begin{eqnarray*}
\fl\log(A) \in [-23.3,-21], \qquad
f_{\rm max} \in [0.001,0.5] \: {\rm Hz},
\\ \fl t_c \in [T_0,T_0+\Delta T] \: {\rm s}, \qquad
\sin(\beta) \in [-1,1], \qquad  \phi \in [0,2\pi], \qquad \psi \in [0,2\pi] \:.
\label{eqn:CSpriors}
\end{eqnarray*}
Here, $\beta$ is the sky latitude, $\phi$ is the sky azimuthal angle, and $\psi$ is the polarization. The data stream was divided into segments for the search and $T_0$ is the start time of the particular data segment being searched and $\Delta T$ is that segment's length. 

For the sine-Gaussian model, we maintained the approach of using uniform prior ranges while making sure that all signals would fall within our specified bounds. The priors used were
\begin{eqnarray*}
\label{eqn:SGpriors}
\fl\log(A) \in [-22.3,-18],
\qquad f_c \in [0.001,0.5] \: {\rm Hz},
\qquad Q \in [1,10] \\
\fl t_c \in [T_0,T_0+\Delta T] \: {\rm s}, \qquad
\sin(\beta) \in [-1,1], \qquad
\phi \in [0,2\pi], \qquad \psi \in [0,2\pi] \:.
\end{eqnarray*}
$T_0$, $\Delta T$, $\beta$, $\phi$, and $\psi$ are the same physical quantities as in the cosmic string model; the prior on $\log(A)$ was modified in order to cover the same range of SNRs for the signals.

\section{Results}
\label{sec:res}
\subsection{MLDC challenge 3.4 search}
\label{sec:mldcres}
Mock LISA Data Challenge 3.4~\cite{mldc} consisted of a data set of $2^{21}$ samples at a cadence of 1s, containing GW burst signals from cosmic string cusps~\cite{CScusps1,CScusps2}. The signals occurred as a Poissonian random process throughout the data set, with an expectation value of five events. To analyse the data, we divided the data stream into segments of 32768s in length, with 2048s of overlap between neighbouring segments. The overlap was chosen to ensure that each of the bursts, which were of typical duration of $\sim1000$s, would be entirely contained in at least one segment. To avoid artefacts in the Fourier domain, we used Welch windowing when performing the Fast Fourier Transform (FFT) of each segment. We also analysed data segments of the same length, offset from the first set by $16384$s, to check no signals had been missed and to verify that the detected signals were not edge artefacts. We also repeated the search with segments of $16384$s and $8192$s in length to verify our results. \MN~was run with $4000$ live points on each segment, using 8 3GHz Intel Woodcrest processors
~in parallel; the algorithm took approximately 5 minutes to run on a single segment when no signal was present, and twice this when one was present. The search of the offset and shorter segments returned all of the same detections. We ran the search on the MLDC training data set to test the algorithm, and then on the challenge data set.

To assess the quality of our solutions we will not only use the recovered parameters for a source, but also the signal-to-noise ratio (SNR) of the recovered solution. The degeneracies in the parameter space mean that a waveform can match the true waveform very well with very different parameter values. The recovered SNR is a measure of how much of a residual would be left in the data stream if the recovered parameters were used to subtract a given source from the data set. If the recovered SNR is close to the true SNR we can say that we have correctly identified the waveform of the source even if the parameters are quite different.

\subsubsection{Training data}
\begin{figure}
\centering
\includegraphics[width=4in]{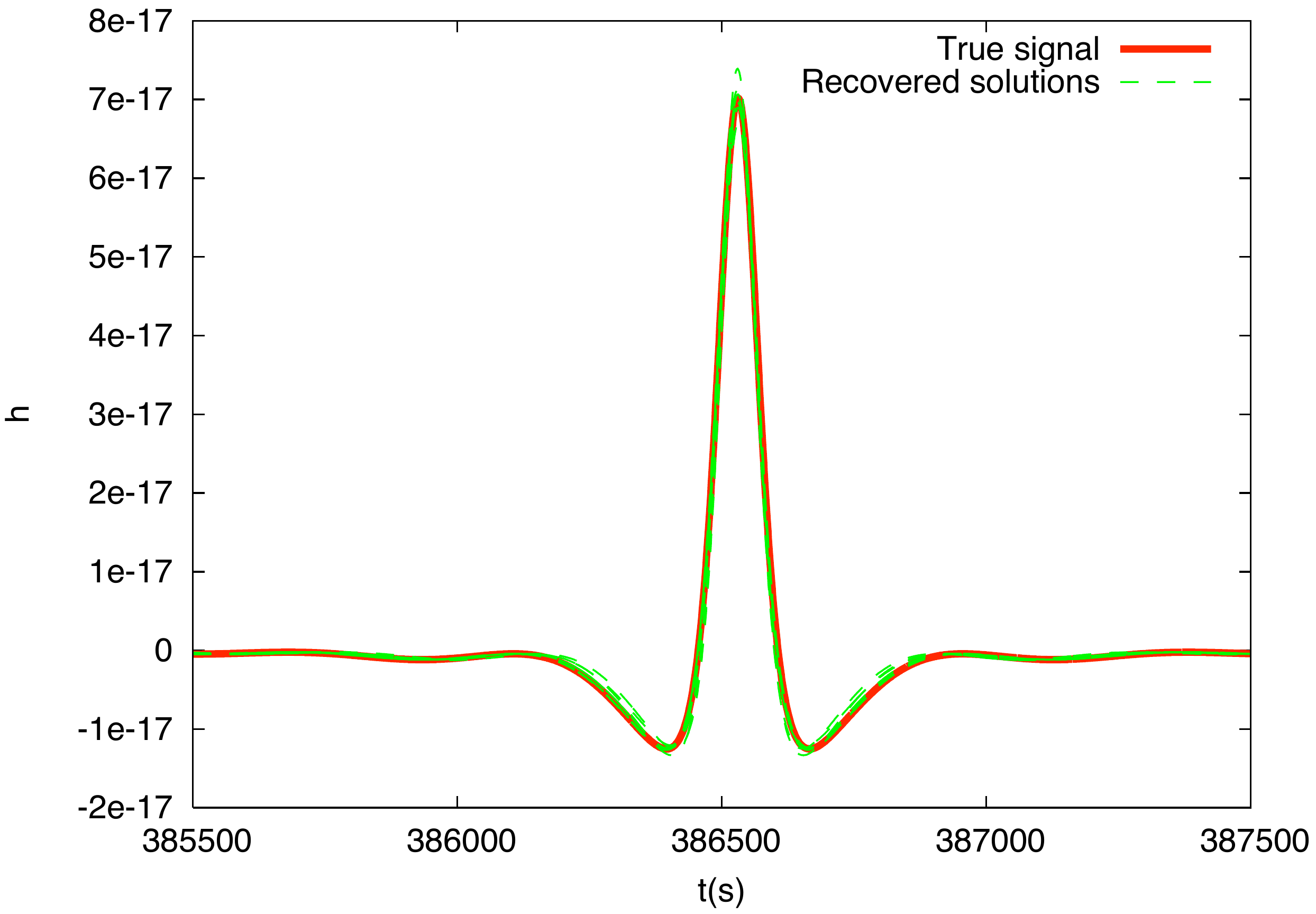}
\caption{True signal and the nine modes of highest evidence identified by the algorithm in our search for the first source in the training data.}
\label{fig:CStrainmodes}
\end{figure}

The MLDC training data set for challenge 3.4 contained five cosmic string bursts. Our search successfully identified the five sources, and recovered posterior probability distributions for their parameters. The posteriors were highly degenerate, containing many modes. These modes corresponded to degeneracies in the waveform parameter space that arise because the burst signals are of short duration compared to the orbital period of the detector. Such degeneracies were noted in~\cite{CornishCS}. We can use the local Bayesian evidence of the modes, as computed by {\sc MultiNest}, to classify the various solutions. In Figure~\ref{fig:CStrainmodes} we show the waveforms corresponding to the nine modes of highest local evidence identified in the search for the first of the signals in the training data set. We also show the true waveform signal for comparison. It is clear that all of the modes that were identified have nearly identical waveforms, so we would not expect to determine uniquely the true source parameters. However, it is also clear that {\sc MultiNest} is successfully finding and identifying the signal.

In Table~\ref{tab:trainpar} we summarise the parameters of the five signals in the training data set and the solutions recovered by {\sc MultiNest}. For simplicity we present only the solution of highest local evidence identified by the algorithm, although in all cases we found multiple other modes. We quote both the maximum likelihood parameters found by {\sc MultiNest} for this mode, and the mean and standard deviation computed from the posterior of that mode. We see that when the maximum frequency, $f_{\rm max}$, is significantly below Nyquist, then we can determine it well, but not when it is close to or above that frequency. This is entirely expected, as this parameter only modifies the signal for $f> f_{\rm max}$ and therefore does not leave a signature in the data set when $f_{\rm max}$ is too large. There is a degeneracy in the polarisation angle corresponding to the shift $\psi \rightarrow \psi+\pi$, and accounting for this the polarisation is moderately well determined, as is the amplitude. The time of coalesence at the detector is extremely well determined, but $t_c$ is the time of coalesence at the Solar System barycentre, and this is not so well determined due to uncertainties in the sky position. The three parameters, $\beta_S, \phi_S, t_c$, show correlated degeneracies. These arise because the ability of LISA to resolve sky position relies on the motion of the detector around the Sun, but over the duration of a typical burst LISA is essentially static. At higher gravitational wave frequencies some of this degeneracy is weakly broken, but it is very severe at low frequency. There is also a well known degeneracy in the LISA response between antipodal sky positions, corresponding to gravitational waves crossing the detector in opposite directions.

The standard deviations of the modes are quite large in all cases, which reflects the existence of these degeneracies. In most cases, the true parameters are within $3$ standard deviations of the means, so the mean and error returned by {\sc MultiNest} adequately reflect the parameters of the underlying signal. One interesting result is that, for source 4, the distribution of $f_{\rm max}$ is as broad as it is in the three cases where this parameter could not be measured (sources 2, 3 and 5), and yet the maximum likelihood value of $0.0118$Hz is very close to the true value of $0.0115$Hz. For this source, the posterior recovered by {\sc MultiNest} was very broad, with a narrow peak in the vicinity of the true value. This suggests that $f_{\rm max}$ may be difficult to constrain even when it has a value significantly below Nyquist.

\begin{table}
\begin{tabular}{|cc|c|c|c|c|c|c|}
\hline \multicolumn{2}{|c|}{Source}&A&$\psi$&$\beta_S$&$\phi_S$&$f_{\rm max}$ (Hz)&$t_c$ (s)\\\hline \multirow{4}{*}{1}&True&$1.98\times10^{-21}$&$0.400$&$-0.249$&$5.59$&$0.00237$&$3.8684\times10^5$\\
&Recovered&$1.80\times10^{-21}$&$3.67$&$-0.577$&$5.83$&$0.00237$&$3.8692\times10^5$\\
&Mean&$1.84\times10^{-21}$&$3.64$&$-0.157$&$5.78$&$0.00238$&$3.8690\times10^5$\\
&Std. Dev.&$\pm0.21\times10^{-21}$&$\pm0.076$&$\pm0.17$&$0.16$&$\pm0.00008$&$\pm0.00042\times10^5$\\
\hline \multirow{4}{*}{2}&True&$1.07\times10^{-21}$&$5.21$&$-1.18$&$4.79$&$1.15$&$1.8892\times10^6$\\
&Recovered&$8.14\times10^{-22}$&$5.61$&$-0.577$&$4.71$&$0.5^{*}$&$1.8891\times10^6$\\
&Mean&$9.21\times10^{-22}$&$4.88$&$-0.553$&$4.88$&$0.24$&$1.8892\times10^6$\\
&Std. Dev.&$\pm0.89\times10^{-22}$&$\pm0.34$&$\pm0.16$&$\pm0.34$&$\pm0.14$&$\pm0.00014\times10^6$\\
\hline \multirow{4}{*}{3}&True&$6.60\times10^{-22}$&$3.98$&$-0.934$&$0.677$&$0.464$&$1.8645\times10^6$\\
&Recovered&$9.96\times10^{-22}$&$1.35$&$-0.703$&$1.55$&$0.5^{*}$&$1.8644\times10^6$\\
&Mean&$8.77\times10^{-21}$&$1.34$&$-0.835$&$1.42$&$0.26$&$1.8644\times10^6$\\
&Std. Dev.&$\pm0.83\times10^{-22}$&$\pm0.19$&$\pm0.14$&$\pm0.22$&$\pm0.14$&$\pm0.000071\times10^6$\\
\hline \multirow{4}{*}{4}&True&$2.65\times10^{-21}$&$3.54$&$0.239$&$1.09$&$0.0115$&$2.0603\times10^6$\\
&Recovered&$2.42\times10^{-21}$&$0.505$&$0.817$&$2.35$&$0.0118$&$2.0598\times10^6$\\
&Mean&$2.31\times10^{-21}$&$0.517$&$0.505$&$2.50$&$0.244$&$2.0597\times10^6$\\
&Std. Dev.&$\pm0.30\times10^{-21}$&$\pm1.3$&$\pm0.20$&$\pm0.10$&$\pm0.15$&$\pm0.000049\times10^6$\\
\hline \multirow{4}{*}{5}&True&$3.42\times10^{-22}$&$1.09$&$-1.03$&$1.16$&$2.28$&$1.4983\times10^6$\\
&Recovered&$5.39\times10^{-22}$&$4.50$&$-0.838$&$1.68$&$0.5^{*}$&$1.4982\times10^6$\\
&Mean&$4.23\times10^{-22}$&$4.04$&$-0.419$&$2.99$&$0.257$&$1.4983\times10^6$\\
&Std. Dev.&$\pm1.1\times10^{-22}$&$\pm1.0$&$\pm0.51$&$\pm2.01$&$\pm0.14$&$\pm0.00024\times10^6$\\\hline
\end{tabular}
\caption{\label{tab:trainpar}Parameters of the five signals present in the training data set, and the parameters recovered by {\sc MultiNest} for the mode of highest local evidence identified by the search. The parameters are amplitude, $A$, polarisation, $\psi$, sky latitude, $\beta_S$, and longitude, $\phi_S$, break frequency, $f_{\rm max}$, and time of coalesence at the Solar System barycentre, $t_c$. In the row labelled ``Recovered'' we quote the maximum likelihood parameters of the mode with highest evidence identified by {\sc MultiNest}. In the rows labelled ``Mean'' and ``Std. Dev.'' we quote the mean and standard deviation in that parameter, as computed from the recovered posterior for the mode of highest evidence. Where the value of $f_{\rm max}$ is marked by a $^{*}$, the distribution in this parameter was very flat, indicating that $f_{\rm max}$ was probably above Nyquist. In these cases, we set $f_{\rm max}$ to the Nyquist frequency of $0.5$Hz for the maximum likelihood parameters, but still quote the actual recovered mean and standard deviation.}
\end{table}

\begin{figure}
\centering
\includegraphics[width=\textwidth]{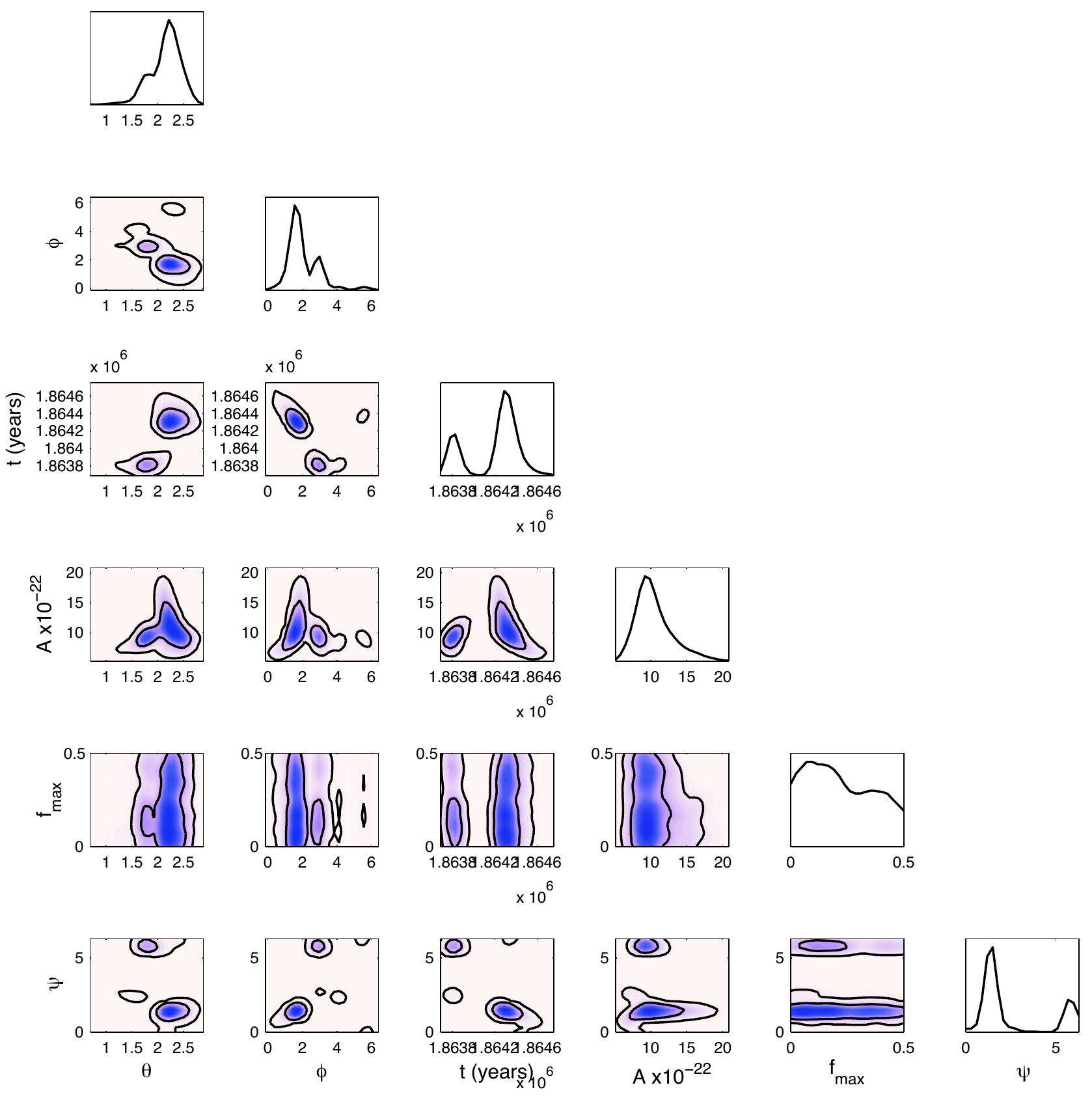}
\caption{Two-dimensional marginalised posteriors as recovered by {\sc MultiNest} in the search for the third of the cosmic string bursts in the MLDC training data set. The parameters, from top-to-bottom and left-to-right, are colatitude, longitude, burst time, burst amplitude, burst break frequency and waveform polarization. At the top of each column we also show the one-dimensional posterior for the column parameter.}
\label{fig:ts3tri}
\end{figure}

The degeneracies in the parameter space are also evident in the marginalised posterior probability distributions, which we can construct from the {\sc MultiNest} live point set evolution. In Figure~\ref{fig:ts3tri} we show 2D and 1D marginalised posterior distributions for the third of the sources present in the training data set. We note that the break frequency, $f_{\rm max}$, is unconstrained as it was only slightly below ($0.464$Hz) the Nyquist frequency of the data set ($0.5$Hz) in this case. The degeneracies in the posteriors for sky position and time of coalesence are also evident. We see two dominant peaks in the sky position posterior, which are points antipodal to one another on the sky.

The existence of these degeneracies in the parameter space suggest that a comparison of recovered and true parameters is not the best way to characterise the quality of a detection, as mentioned above. An alternative approach is to compute the signal-to-noise ratio of the recovered source and compare that to the SNR of the injected source. This was possible using evaluation tools available in the {\sc lisatools} code library\footnote{http://code.google.com/p/lisatools/}. In Table~\ref{tab:trainolps} we show the SNRs in the $A$ and $E$ channels for each of the five training sources. We give results for the recovered mode of highest local evidence, plus the corresponding result using the true source parameters. We see that, in all cases, the solution found by {\sc MultiNest} recovered almost all of the SNR of the true source. If the burst source was subtracted from the data set using the recovered parameters, the residual in the data set would be at the level of instrumental noise fluctuations. We can therefore say that we correctly identified all of the sources.

\begin{table}
\begin{tabular}{|cc|c|c|c|c|c|}
\hline &&\multicolumn{5}{|c|}{Source}\\
Channel&&1&2&3&4&5\\\hline
\multirow{2}{*}{A}&True&52.4&12.9&16.1&12.6&4.34\\
&Recovered&52.4&12.7&16.5&12.6&4.19\\\hline
\multirow{2}{*}{E}&True&17.1&17.7&26.8&76.6&13.6\\
&Recovered&17.1&17.6&26.7&76.6&13.6\\\hline
\end{tabular}
\caption{\label{tab:trainolps}SNRs in the $A$ and $E$ channels for the five sources present in the training data, as computed using the evaluation scripts in {\sc lisatools}. We quote results using the true source parameters, labelled ``True'', and using the parameters of the mode identified by {\sc MultiNest} of highest local evidence, labelled ``Recovered''.}
\end{table}

\subsubsection{Challenge data}
Three signals were found in the challenge data set at $t_c\sim 6\times10^5$s, $t_c\sim 1.07\times 10^6$s and $t_c \sim 1.6\times 10^6$s and this was the correct number of signals. We label these as source candidates 3, 2 and 1 respectively for consistency with the MLDC key file. For source candidates 1 and 3, {\sc MultiNest} returned a flat distribution in $f_{\rm max}$, thereby leading us to conclude that the actual maximum was above the Nyquist frequency of 0.5 Hz. We identified several modes in the posterior in each case, and used the evidence to characterise these as for the training data. We decided to submit the two modes of highest evidence, as in the analysis of the training data these generally corresponded to two antipodal sky solutions. However, for source candidate 3 there was a third mode of almost equal local evidence, and so we submitted that mode as well. The middle source, 2, was found to have a break frequency of 0.0011 Hz, very close to the minimum value in the prior range of 0.001 Hz. At this low frequency, LISA is not able to resolve the sky position for a burst source. As expected, we found very broad posteriors for all parameters other than the maximum frequency. As the posterior was not well separated into modes, we chose to submit only one set of parameters in this case, which we took to be the maximum a-posteriori parameters.

In Table~\ref{tab:challpar} we list the true parameters for the three sources in the challenge data set\footnote{Available at {\tt http://www.tapir.caltech.edu/$\sim$mldc/results3/}}, and the parameters recovered by {\sc MultiNest}. For simplicity we only include the parameters for the mode of highest local evidence, and not all of the modes submitted. The agreement between the true and recovered parameters is broadly consistent with our expectations from the analysis of the training data, and once again the true parameters lie within a few standard deviations of the means in most cases\footnote{For source 1, $t_c$ is quite far from the true value by this measure. However, this is due to the correlated degeneracies between $t_c$, $\beta_S$ and $\phi_S$. The true time of coalesence was consistent with the full recovered posterior distribution.}. One surprise is that for source 1, the true value of $f_{\rm max}$ is considerably below Nyquist, while our analysis recovered a flat distribution indicative of a value above Nyquist. We have carried out further checks, but there is no indication that the data favours a particular value of $f_{\rm max}$ in this case. This is similar to what we observed for source 4 in the training data, which had a similar value of $f_{\rm max}$. However, in that case there was a narrow peak in the posterior in the vicinity of the true value, but no such peak was evident here.

\begin{table}
\begin{tabular}{|cc|c|c|c|c|c|c|}
\hline \multicolumn{2}{|c|}{Source}&A&$\psi$&$\beta_S$&$\phi_S$&$f_{\rm max}$ (Hz)&$t_c$ (s)\\\hline \multirow{4}{*}{1}&True&$8.66\times10^{-22}$&$3.32$&$0.556$&$3.71$&$0.0296$&$1.6022\times10^6$\\
&Recovered&$1.65\times10^{-21}$&$2.82$&$0.349$&$6.01$&$0.5^{*}$&$1.6030\times10^6$\\
&Mean&$1.13\times10^{-21}$&$2.70$&$-0.141$&$4.38$&$0.27$&$1.6030\times10^6$\\
&Std. Dev.&$\pm0.22\times10^{-21}$&$\pm0.14$&$\pm0.30$&$\pm2.6$&$\pm0.14$&$\pm0.000071\times10^6$\\\hline
\hline \multirow{4}{*}{2}&True&$2.79\times10^{-21}$&$5.12$&$-0.444$&$3.17$&$0.00108$&$1.0727\times10^6$\\
&Recovered&$2.97\times10^{-21}$&$0.271$&$-0.893$&$5.12$&$0.00108$&$1.0732\times10^6$\\ 
&Mean&$2.75\times10^{-21}$&$0.91$&$-0.00804$&$5.31$&$0.00108$&$1.0733\times10^6$\\ &Std. Dev.&$\pm0.63\times10^{-21}$&$\pm0.28$&$\pm0.52$&$\pm0.46$&$\pm0.000048$&$\pm0.00019\times10^6$\\\hline
\hline \multirow{4}{*}{3}&True&$8.54\times10^{-22}$&$4.66$&$-0.800$&$0.217$&$6.15$&$6.0000\times10^5$\\
&Recovered&$1.14\times10^{-21}$&$0.927$&$-0.562$&$5.45$&$0.5^{*}$&$5.9991\times10^5$\\
&Mean&$1.16\times10^{-21}$&$0.962$&$-0.488$&$5.48$&$0.24$&$5.9992\times10^5$\\
&Std. Dev.&$\pm0.18\times10^{-21}$&$\pm0.16$&$\pm0.16$&$\pm0.18$&$\pm0.14$&$\pm0.00076\times10^5$\\\hline
\end{tabular}
\caption{\label{tab:challpar}As Table~\ref{tab:trainpar}, but now for the three signals in the challenge data set.}
\end{table}

\begin{figure}
\begin{center}
\begin{tabular}{c}
\includegraphics[width=0.6\textwidth]{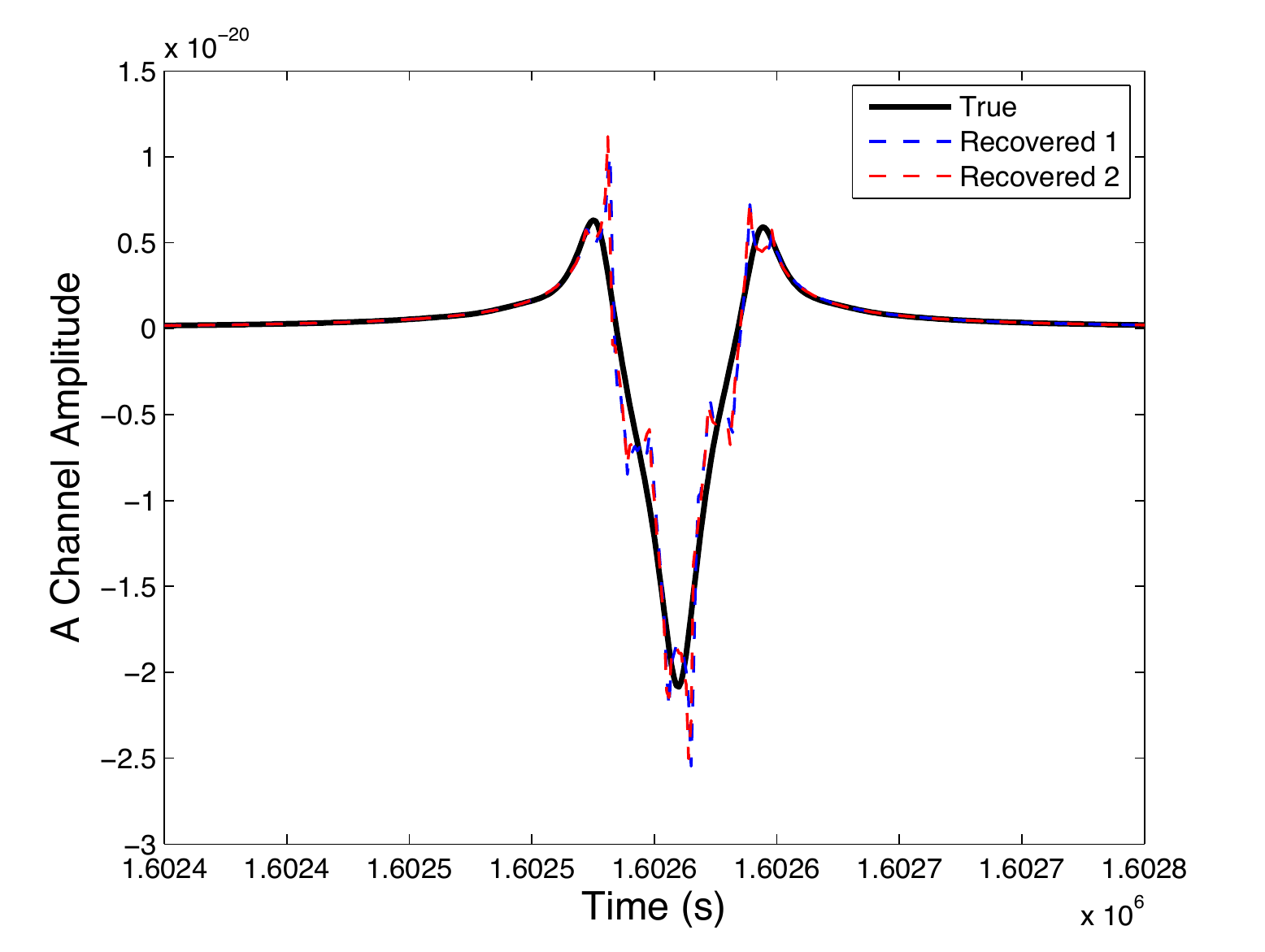}\\ Source 1\\
\includegraphics[width=0.6\textwidth]{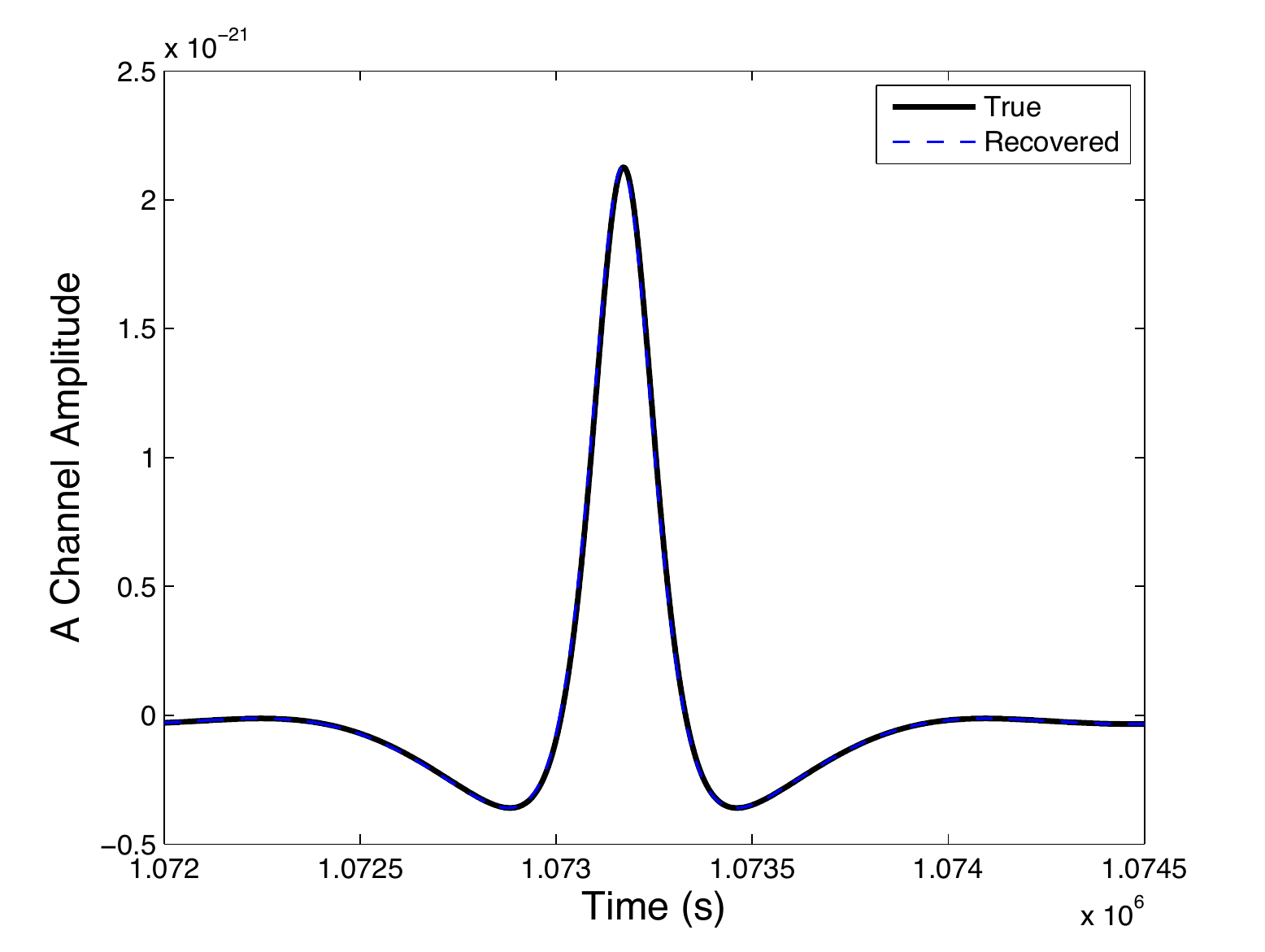}\\ Source 2\\
\includegraphics[width=0.6\textwidth]{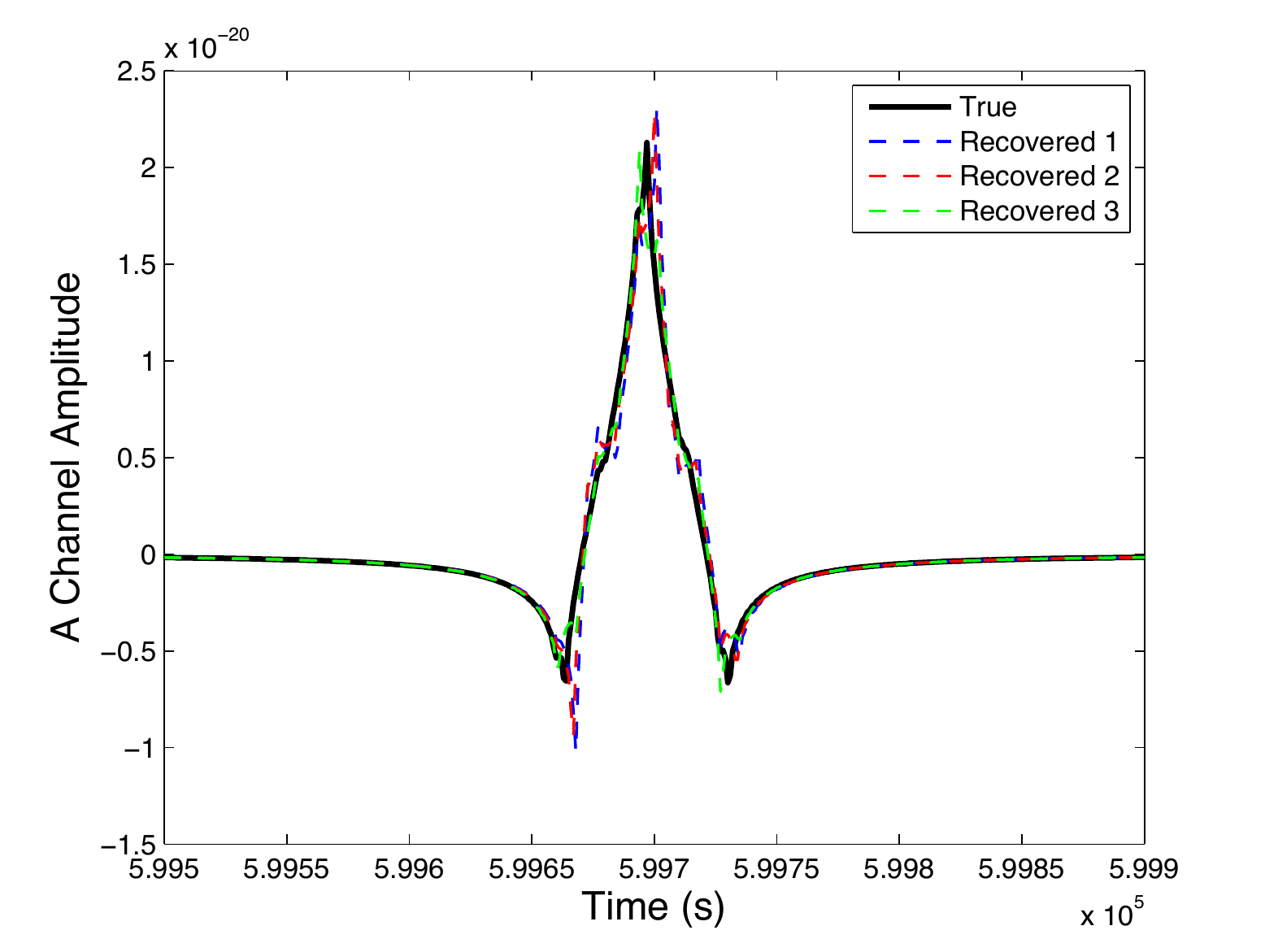}\\ Source 3\\
\end{tabular}
\end{center}
\caption{\label{fig:CamVsTrue}Plots comparing the true waveforms with the waveforms for the various modes recovered by {\sc MultiNest} and included in our submission for the analysis of the challenge data set.}
\end{figure}

In Figure~\ref{fig:CamVsTrue} we show the true waveforms, in the $A$ channel, for the three sources present in the data set, and overlay them with the waveforms  corresponding to each of the modes that we included in our submission. We see clearly that, in all three cases, the recovered waveforms reproduce the true waveforms well, with small differences that arise from noise fluctuations in the detector. As for the training data, we expect the recovered SNR to be a better measure of the quality of the solution than the values of the recovered parameters. In Table~\ref{tab:challolps} we list the true SNRs in the $A$ and $E$ channels, and the SNRs recovered by each of the solutions included in the submission. As we found in the case of the training data, we have clearly recovered all of the SNR of the true signal, and the residual after subtraction would be at the level of instrumental noise. We thus regard the search as a success.

\begin{table}
\begin{tabular}{|c|c|c|c|ccc|}
\hline
&&&&\multicolumn{3}{c|}{Recovered SNR}\\
Source&$t_c$(s)&Channel&True SNR&Mode 1&Mode 2&Mode 3\\\hline
1&$1.6\times10^6$&\begin{tabular}{l}A\\ E\end{tabular}&\begin{tabular}{l}41.0\\ 14.5\end{tabular}&\begin{tabular}{l}41.2\\ 14.5\end{tabular}&\begin{tabular}{l}41.0\\ 14.6\end{tabular}&N/A\\\hline
2&$1.1\times10^6$&\begin{tabular}{l}A\\ E\end{tabular}&\begin{tabular}{l}30.7\\ 13.9\end{tabular}&\begin{tabular}{l}30.7\\ 13.9\end{tabular}&N/A&N/A \\\hline
3&$6\times10^5$&\begin{tabular}{l}A\\ E\end{tabular}&\begin{tabular}{l}18.8\\ 36.9\end{tabular}&\begin{tabular}{l}18.9\\ 36.7\end{tabular}&\begin{tabular}{l}18.5\\ 37.1\end{tabular}&\begin{tabular}{l}18.4\\ 36.8\end{tabular} \\\hline
\end{tabular}
\caption{\label{tab:challolps}Signal-to-noise ratios for the three sources in the challenge data set. We show the true SNRs in the $A$ and $E$ channels, plus the SNRs for all of the modes we submitted for analysis.}
\end{table}

\begin{figure}
\centering
\includegraphics[width=\textwidth]{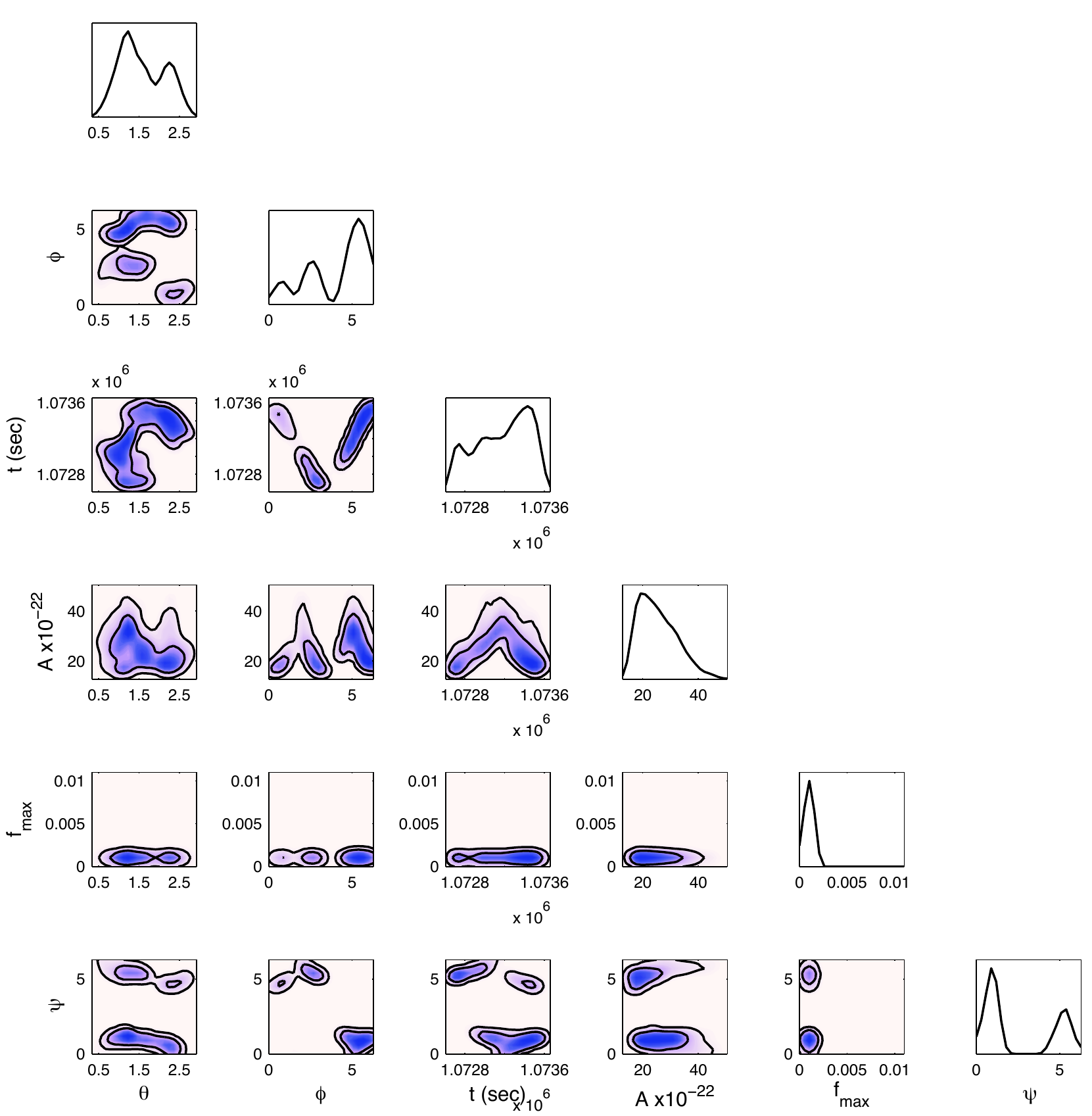}
\caption{As Figure~\ref{fig:ts3tri}, but now for the second of the sources identified in the blind challenge data set.}
\label{fig:bs2tri}
\end{figure}

In Figure~\ref{fig:bs2tri} we show 1D and 2D marginalised posterior probability distributions for the second of the sources present in the challenge data set. Comparing to the posteriors for the training source shown in Figure~\ref{fig:ts3tri}, we see that, in this case, $f_{\rm max}$ can be measured very well, since it is far below Nyquist, but similar degeneracies in sky position and time of coalesence are clearly evident. In this case the sky position degeneracies are particularly severe since the source is at very low frequency.

\subsection{Model selection using Bayesian evidence}
The MLDC search demonstrated that {\sc MultiNest} is able to correctly identify and characterize bursts from cosmic string cusps in LISA data. However, cosmic string cusps produce a very particular waveform signature. LISA might also detect bursts of gravitational radiation from other sources, such as supernovae of hypermassive objects, or even unmodelled sources. A search of the LISA data using a cosmic string cusp model could conceivably detect these other types of burst. In order to make scientific statements it is important to be able to say what the most likely origin for an observed burst could be. The Bayesian evidence provides a tool for carrying out model selection of this type, although it relies on having two alternative hypotheses to compare. As an alternative to a cosmic string model, we consider a sine-Gaussian burst, which is a generic burst waveform commonly used in burst analysis for ground-based gravitational wave detectors~\cite{LSCUsingSG}. We can use the evidence ratio to compare these two models when analysing a data stream containing a cosmic string burst or containing a sine-Gaussian burst. There are two stages to this analysis --- 1) determining the signal-to-noise ratio at which we begin to be able to detect a signal of each type in a LISA data set; 2) determining the SNR at which the Bayesian evidence begins to favour the true model over the alternative.

\subsubsection{Detection threshold}
To determine the SNR needed for detection of a burst source with the \MN~algorithm, we generated a sequence of data sets containing a particular source and Gaussian instrumental noise generated using the theoretical noise spectral densities given in Eq.~(\ref{specdens}), but varying the burst amplitude between the data sets to give a range of SNRs between $5$ and $15$. Note that here and in the following we quote total SNRs for the $A$ and $E$ channels combined. Each data set was searched with  \MN~and the global log-evidence of the data set, as computed by \MN,~was recorded. This was repeated for $8$ different cosmic string sources and $9$ different sine-Gaussian sources. The parameters for the cosmic string sources were taken from the parameters of the five training and three blind sources injected in the MLDC data sets, as these covered the parameter space of possible signals nicely. The important parameters of the sine-Gaussian model are the frequency and the burst width, so we considered three different values of each, and constructed waveforms for the nine possible combinations, while choosing a random value for the sky position and waveform polarisation in each case.

\begin{figure}
\centering
\includegraphics[width=4in]{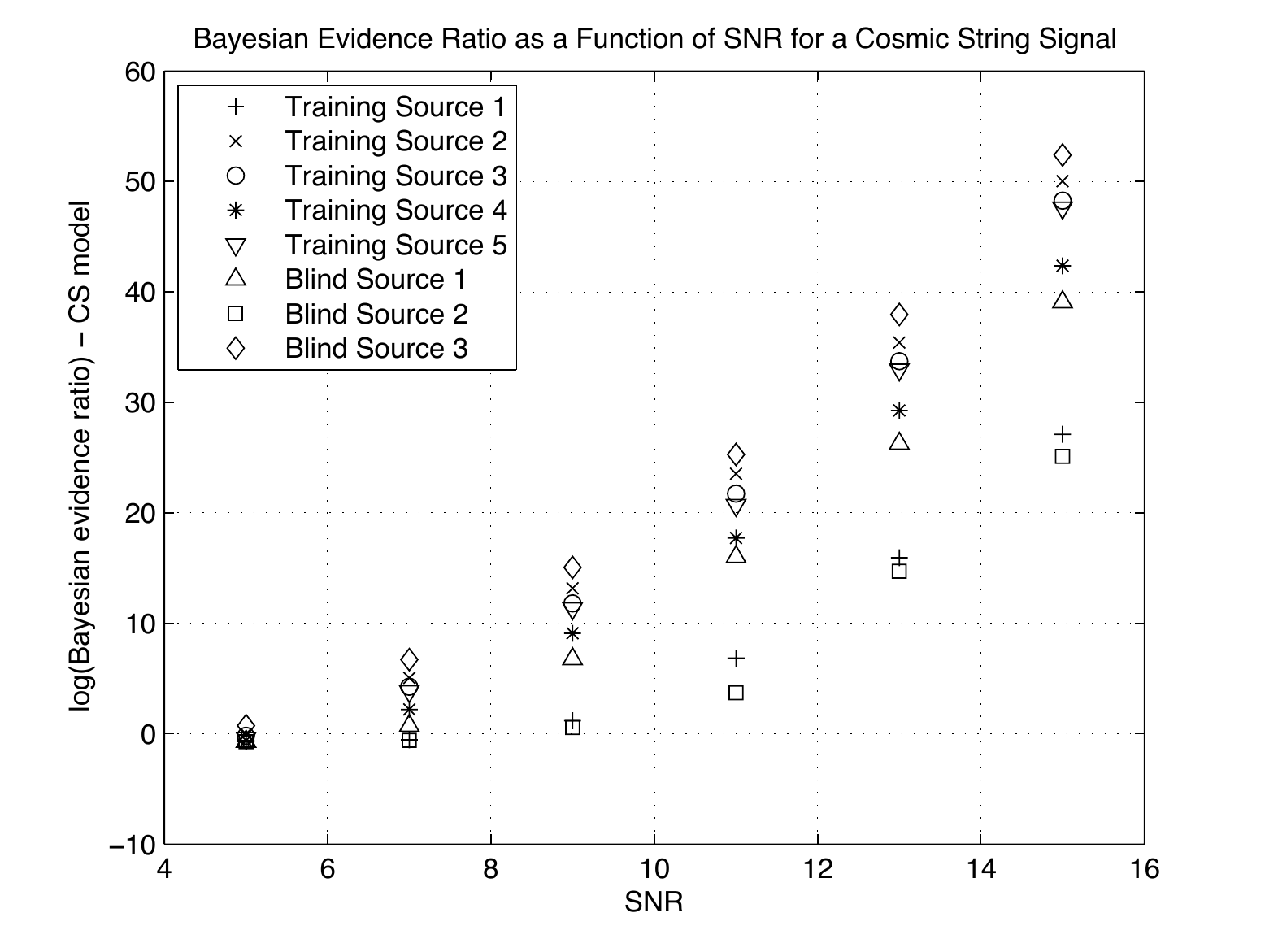}
\caption{The Bayesian log-evidence for the data set as a function of the SNR for eight different cosmic string cusp signals. These signals had the same parameters, other than amplitude, as the various sources used in the training and blind data sets of the MLDC. The labels in the key are consistent with labels in Tables~\ref{tab:trainpar} and~\ref{tab:challpar}.}
\label{fig:varSNRCS}
\end{figure}

\begin{figure}
\centering
\includegraphics[width=4in]{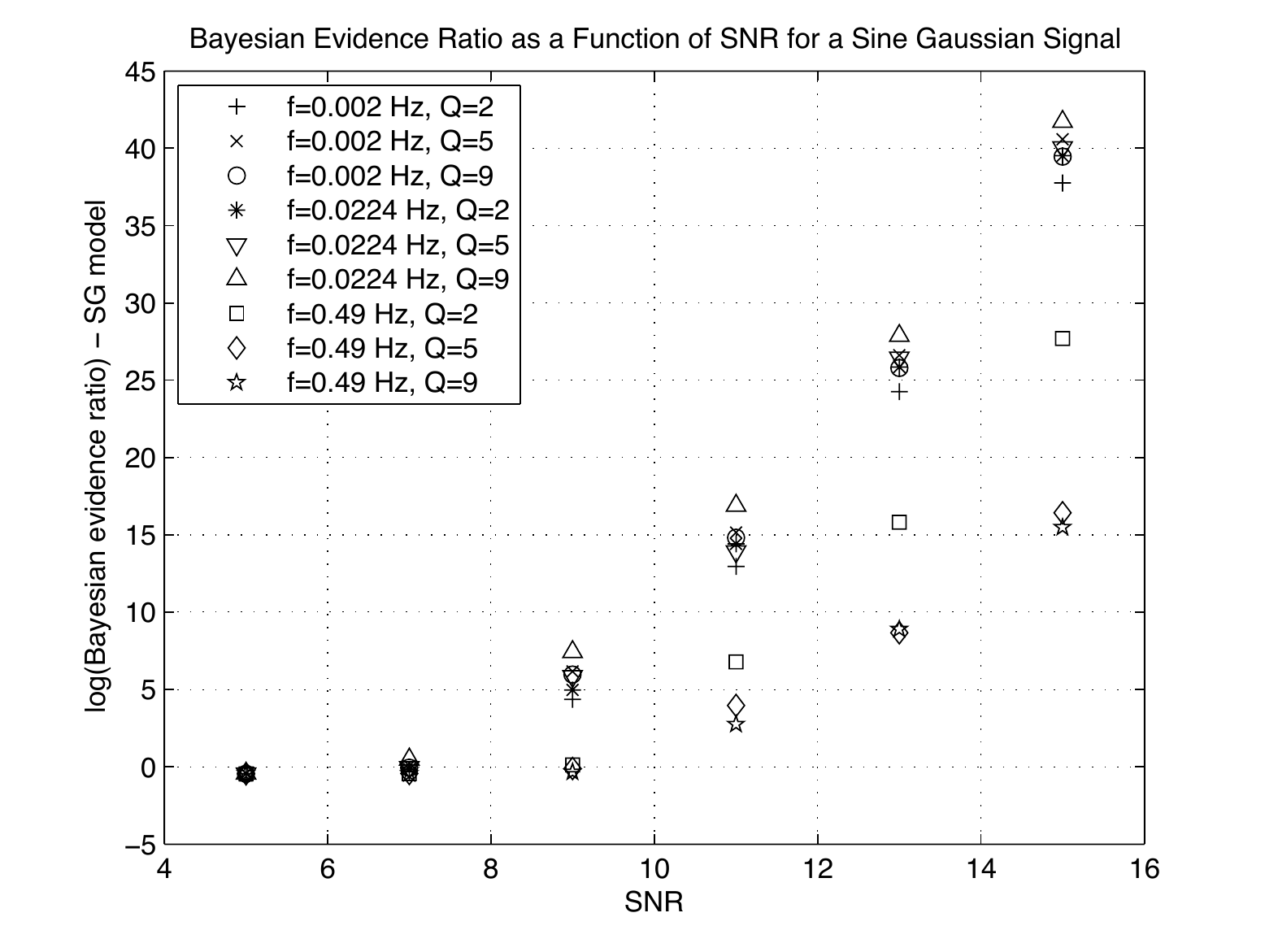}
\caption{As Fig.~\ref{fig:varSNRCS}, but now showing results for nine different sine-Gaussian signals, with frequency, $f$, and width, $Q$, as shown.}
\label{fig:varSNRSG}
\end{figure}

In Figure~\ref{fig:varSNRCS} we show the global log-evidence of the data as a function of the SNR of the injected signal for the cosmic string burst sources. Figure~\ref{fig:varSNRSG} shows the corresponding results for the sine-Gaussian bursts. The detection threshold for cosmic string bursts appears to be at an SNR of $7$--$11$ depending on the source parameters. Most of the sources have a significant log-evidence (greater than 3) at an SNR of 7, but the SNR required for detection is higher for the first training source and the second blind source. These are the two sources with the lowest values of the break frequency $f_{\rm max}$, which makes the waveforms much smoother and simpler in the time domain, as indicated in Fig.~\ref{fig:CStrainmodes}. This may explain why it is more difficult to distinguish them from noise.
The MLDC data sets had a prior for the source SNR, in a {\em single} Michelson channel, of $\rho \in [10,100]$, so these results suggest we would be able to detect any source drawn from the MLDC prior.

The sine-Gaussian bursts require a slightly higher SNR for detection, of $8$--$11$, but this increase in threshold is only of the order of $1$ in SNR. The sine-Gaussian signals are in general much simpler in form than the cosmic strings and therefore it is perhaps unsurprising that it requires a higher SNR to distinguish them from noise. In this case, it was the sources with highest frequency, $f_{c} = 0.49$Hz, that were most difficult to detect. However, since the Nyquist frequency of the data sets was $0.5$Hz, the difficulty of detection may have arisen because the signal was partially out of band.

\subsubsection{Model selection}
\begin{figure}
\centering
\includegraphics[width=4in]{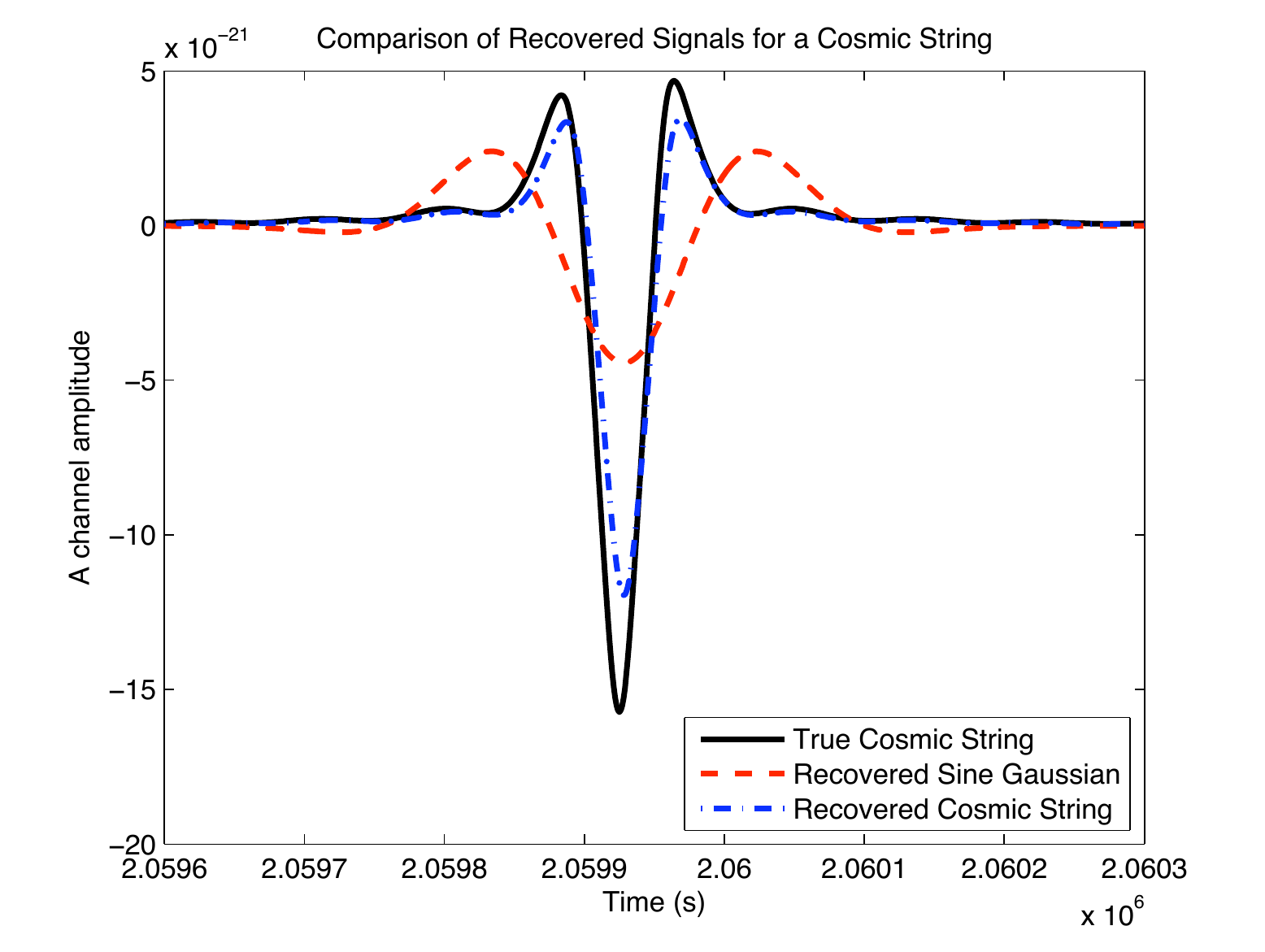}
\caption{Typical example of confusion when searching a cosmic string data set with the wrong model. The plot shows a comparison of the injected cosmic string signal to the best-fit signals found by \MN~using the cosmic string model as templates and using the sine-Gaussian model as templates.}
\label{fig:cosstringrec}
\end{figure}

\begin{figure}
\centering
\includegraphics[width=4in]{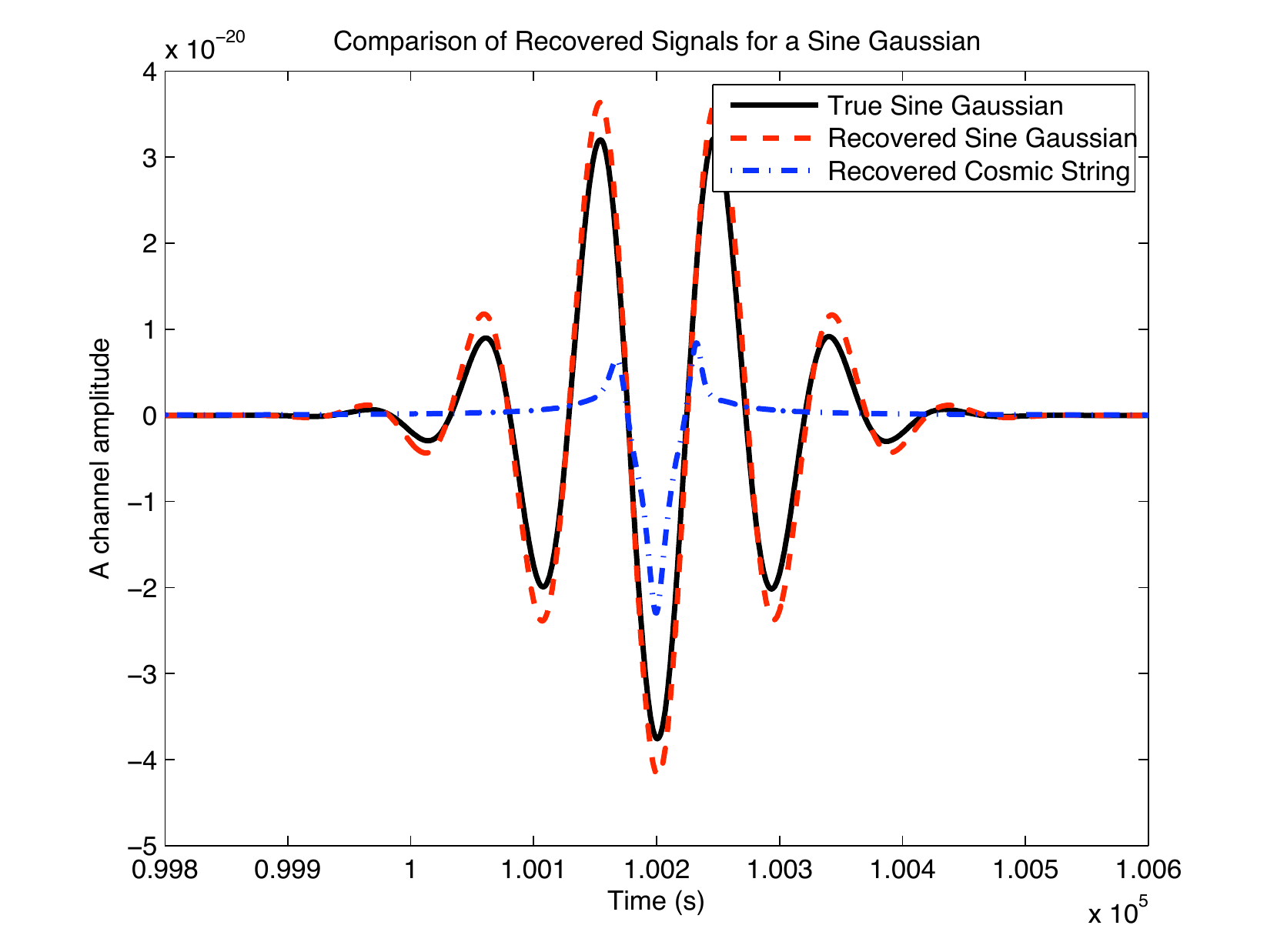}
\caption{As Fig.~\ref{fig:cosstringrec}, but we now show a typical confusion example for searches of the sine-Gaussian data sets. We compare the injected sine-Gaussian signal to the best-fit signals recovered by \MN~using both the cosmic string and the sine Gaussian models.}
\label{fig:sinegaussrec}
\end{figure}

To explore model selection, we used \MN~to search the same data sets described above, but now using the alternative model, i.e., we searched the cosmic string data sets using the sine-Gaussian likelihood and vice versa. For sufficiently high SNR, in both cases the alternative model was able to successfully detect a signal in the data set. Typically, the best-fit sine-Gaussian signal to a cosmic string source has low $Q$ and a frequency that matches the two lobes of the cosmic string burst. The parameter $Q$ sets the number of cycles in the sine-Gaussian wave packet, and so a sine-Gaussain with $Q\sim 2$ most closely resembles a cosmic string event, which typically has two cycles. This is illustrated for a typical case in Fig.~\ref{fig:cosstringrec}. Similarly, the best-fit cosmic string source to a sine-Gaussian signal matches the central two peaks of the sine-Gaussian waveform as well as possible. A typical case is shown in Fig.~\ref{fig:sinegaussrec}.

\begin{figure}
\centering
\includegraphics[width=4in]{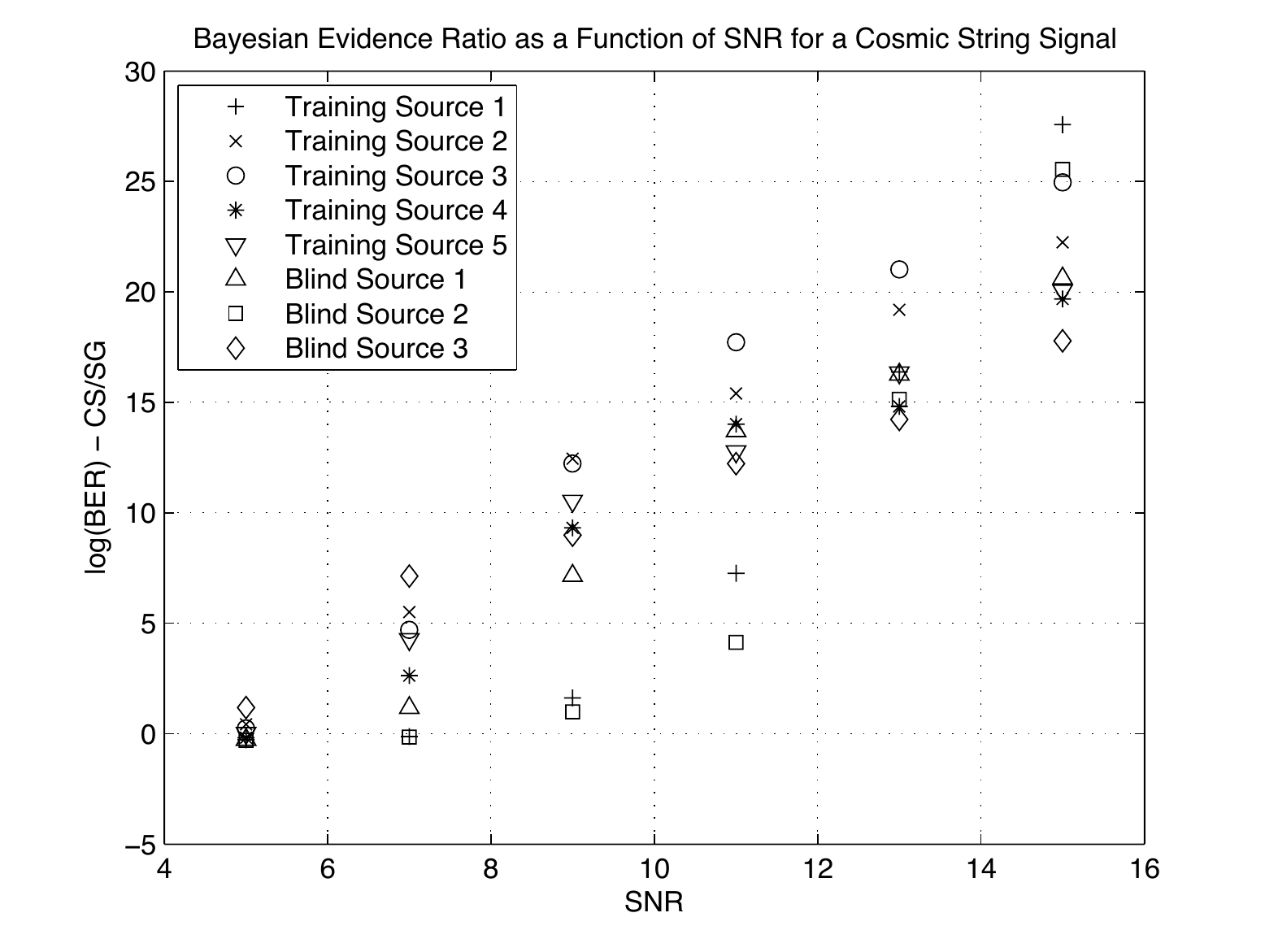}
\caption{The ratio of the Bayesian evidence for the cosmic string model to that of the sine-Gaussian model when searching a data set containing a cosmic string burst source. We show the Bayesian evidence ratio as a function of signal SNR for a variety of different cosmic string sources.}
\label{fig:compmodCS}
\end{figure}

From the results of these searches it is possible to construct the evidence ratio of the two models for each of the data sets. In Fig.~\ref{fig:compmodCS} we show the ratio of the Bayesian evidence for the cosmic string model to that of the sine-Gaussian model when searching the cosmic string data sets. We see that the evidence ratio starts to significantly favour the true model, i.e., the cosmic string, at an injected SNR of $\sim 7$, which is the point at which we first start to be able to detect the cosmic string burst at all. For the two low frequency sources, training source 1 and blind source 2, the evidence ratio only starts to favour the true model at SNR$\sim11$, but again this is the point at which we are first able to detect the source. We conclude that when a cosmic string burst is loud enough to be detected, then the evidence clearly favours the interpretation of the event as a cosmic string burst.

\begin{figure}
\centering
\includegraphics[width=4in]{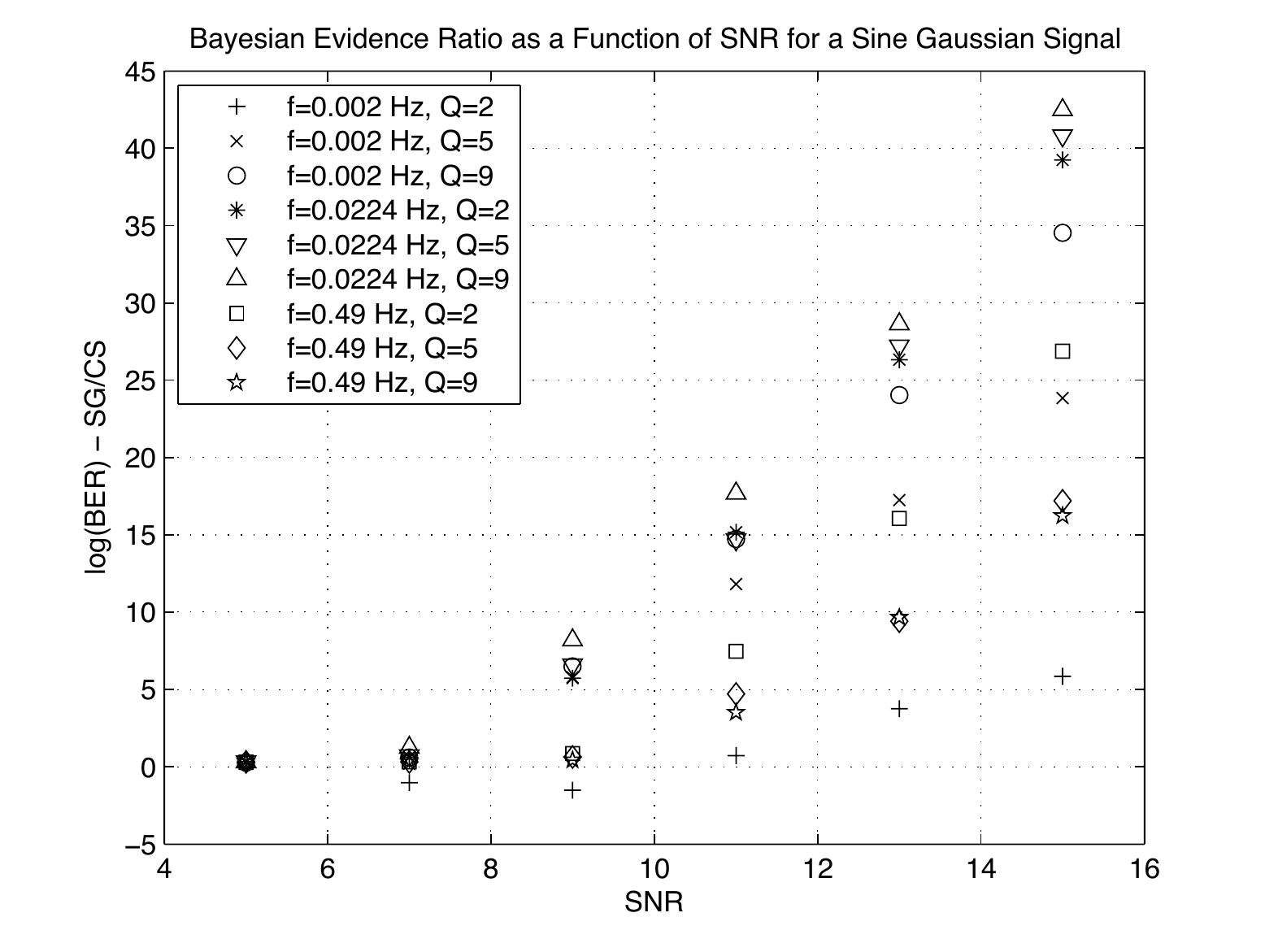}
\caption{As Fig.~\ref{fig:compmodCS}, but now showing the ratio of the Bayesian evidence for the sine-Gaussian model to that of the cosmic string model when searching a data set containing a sine-Gaussian burst source.}
\label{fig:compmodSG}
\end{figure}

In Fig.~\ref{fig:compmodSG} we show the ratio of the evidence of the sine-Gaussian model to that of the cosmic string model when searching the data sets containing sine-Gaussian signals. The conclusions are broadly the same as for the cosmic string case. We require a slightly higher SNR in order to correctly choose the sine-Gaussian model, but this just reflects the fact that we need a somewhat higher SNR to detect the sine-Gaussians in the first place. The only case for which the evidence ratio does not begin to favour the sine-Gaussian model at the point where the source becomes detectable is the case with $f=0.002$Hz and $Q=2$. This is a sine-Gaussian signal with only two smooth oscillations, and so it does look rather like a low frequency cosmic string event. Even in that case, the evidence begins clearly to favour the correct model for SNRs of $\sim13$ and higher.

\section{Discussion}
\label{sec:discuss}
We have considered the use of the multi-modal nested sampling algorithm \MN~for detection and characterisation of cosmic string burst sources in LISA data. As a search tool, the algorithm was able successfully to find the three cosmic string bursts that were present in the MLDC challenge data set. These sources, and the five sources in the MLDC training data, were correctly identified in the sense that the full signal-to-noise ratio of the injected source was recovered, and a posterior distribution for the parameters obtained. The maximum likelihood and maximum a-posteriori parameters were not particularly close to the true parameters of the injected signals, but this was a consequence of the intrinsic degeneracies in the cosmic string model parameter space and in all cases the true parameters were consistent with the recovered posterior distributions.

In controlled studies, we found that the SNR threshold required for detection of the cosmic string bursts was $\sim 7$--$11$, depending on the burst parameters. Bursts with a low break-frequency require a higher SNR to detect than those with high break frequencies. We also explored the detection of sine-Gaussian bursts and in that case the SNR required for detection was slightly higher, being typically $8$--$11$, with sources having frequency close to Nyquist being more difficult to detect.

\MN~is designed to evaluate the evidence of the data under a certain hypothesis, and this can be used to compare possible models for the burst sources. LISA may detect bursts from several different sources, and it is important for scientific interpretation that the nature of the burst be correctly identified. We used the Bayesian evidence as a tool to choose between two different models for a LISA burst source --- the cosmic string model and the sine-Gaussian model, which was chosen to represent a generic burst. The Bayesian evidence works very well as a discriminator between these two models. The evidence ratio begins to clearly favour the correct model over the alternative at the same SNR that the sources become loud enough to detect in the first place.

The usefulness of \MN~as a search tool in this problem is a further illustration of the potential utility of this algorithm for LISA data analysis, as previously demonstrated in a search for non-spinning SMBH binaries~\cite{MNnospin}. Other algorithms based on Markov Chain Monte Carlo techniques have also been applied to the search for cosmic strings~\cite{CornishCS}. Both approaches performed equally well as search tools in the last round of the MLDC. We are now exploring the application of \MN~to searches for other LISA sources, including spinning massive black hole binaries and extreme-mass-ratio inspirals. \MN~was not designed primarily as a search algorithm, but as a tool for evidence evaluation and this work has demonstrated the utility of the Bayesian evidence as a tool for model selection in a LISA context. Other problems where the evidence ratio approach could be applied include choosing between relativity and alternative theories of gravity as explanations for the gravitational waves observed by LISA, or choosing between different models for a gravitational wave background present in the LISA data set. The Bayesian evidence was previously used in a LIGO context as a tool to choose between alternative theories of gravity~\cite{Veitch:2008wd} and in a LISA context to distinguish a data set containing a source from one containing purely instrumental noise~\cite{littenberg09}. \MN~provides a more efficient way to compute the evidence and this should be explored in more detail in the future.

In the context of interpretation of LISA burst events, what we have considered here is only part of the picture. We have shown that we are able to correctly choose between two particular models for a burst, and this can easily be extended to include other burst models. However, LISA might also detect bursts from unmodelled sources. In that case, algorithms such as \MN~which rely on matched filtering would find the best fit parameters within the model space, but a higher intrinsic SNR of the source would be required for detection. In such a situation, we would like to be able to say that the source was probably not from a model of particular type, e.g., not a cosmic string burst. There are several clues which would provide an indication that this was the case. The sine-Gaussian model is sufficiently generic that we would expect it, in general, to provide a better match to unmodelled bursts than the cosmic string model, which has a very specific form. Therefore, we could say that if the evidence ratio favoured the cosmic string model over the sine-Gaussian model it was highly likely that the burst was in fact a cosmic string and not something else. Similarly, if we found that several of the alternative models had almost equal evidence, but the SNR was quite high, it would be indicative that the burst was not described by any of the models. We have seen that at relatively moderate SNRs, when the signal is described by one of the models, the evidence clearly favours the true model over an alternative. If we found that two models gave almost equally good descriptions of the source, it would suggest that the burst was not fully described by either of them. A third clue would come from the shape of the posterior for the source parameters. The cosmic string waveform space contains many degeneracies, but these can be characterised theoretically for a given choice of source parameters. If the signal was not from a cosmic string, we might find that the structure of the posterior was modified. Finally, some techniques have been developed for the Bayesian reconstruction of generic bursts~\cite{roever09} which could also be applied in a LISA context. The usefulness of these various approaches can be explored further by analysing data sets into which numerical supernovae burst waveforms have been injected. While the necessary mass of the progenitor is probably unphysically high for a supernova to produce a burst in the LISA frequency band, such waveforms provide examples of unmodelled burst signals on which to test analysis techniques. The final LISA analysis will employ a family of burst models to characterize any detected events. The work described here demonstrates that the Bayesian evidence will be a useful tool for choosing between such models, and \MN~is a useful tool for computing those evidences.

\section*{Acknowledgements}

This work was performed using the Darwin Supercomputer of the University of Cambridge High Performance Computing Service ({\tt http://www.hpc.cam.ac.uk/}), provided by Dell Inc. using Strategic Research Infrastructure Funding from the Higher Education Funding Council for England and the authors would like to thank Dr. Stuart Rankin for computational assistance. FF is supported by the Trinity Hall Research Fellowship. JG's work is supported by the Royal Society. PG is supported by the Gates Cambridge Trust.

\end{document}